\documentclass[final,3p,times,12pt]{elsarticle}

\usepackage{amssymb,amsmath}
\providecommand{\kwd}[1]{#1\quad}

\DeclareMathOperator*{\argmax}{arg\,max}

\usepackage{setspace,url}
\usepackage{xcolor}
\usepackage{soul}
\usepackage{amsmath,graphicx}
\usepackage{graphics,comment}
\usepackage{tikz}
\usepackage{tikz}
\usetikzlibrary{arrows,patterns}
\usetikzlibrary{positioning}
\usepackage{lineno,hyperref}
\newcommand{\C}[1]{} 
\journal{Journal of \LaTeX\ Templates}

\begin{document}

\begin{frontmatter}

\title{Single frequency filtering based multi-speaker direction of arrival estimation from stereo recordings}

\author{Sushmita Thakallapalli$^1$, Sudarsana Reddy Kadiri$^2$, Nilesh Madhu$^3$, and Suryakanth V Gangashetty$^1$}

\address{$^1$Speech Processing Laboratory, International Institute of Information Technology, Hyderabad, India\\$^2$Signal Analysis and Interpretation Laboratory, University of Southern California, Los Angeles, USA\\$^3$IDLab, Dept. Electronics \& Information Systems, 
Ghent University - imec, Belgium\\}

\begin{abstract} 
Robust estimation of direction of arrival (DoA) of a target source from noisy and reverberant microphone signals is a challenging problem. Conventional DoA estimators such as Generalized cross correlation (GCC) and its variants work in the short-time Fourier transform (STFT) domain on the spectral features, which represent mostly characteristics of vocal tract system during speech production. A few recent estimators operate in a time-frequency (TF) domain called single frequency filtering (SFF). In this domain, the complex outputs at several frequencies are obtained by filtering the speech signals through a single pole filter and the signals are represented with good spectral resolution of harmonics and good temporal resolution of excitation source features such as impulse-like excitations (corresponding to the epoch locations). Such simultaneous good spectral and temporal resolution is not evident in the STFT domain. The SFF-based estimators exploit either spectral or excitation source features of speech production. A study has empirically verified that the excitation source features are more robust to degradations like noise and reverberation than the spectral features, motivating us to further explore methods to improve the existing SFF-based DoA estimators, which exploit the characteristics of excitation source during speech production. In the proposed estimator, the envelopes at several frequencies of SFF outputs are correlated with the corresponding envelopes at the other channel by PHAse Transform (PHAT) weighted GCC. Further, since no existing studies evaluate and compare the SFF-based with GCC-based estimators,  comprehensive evaluations (in terms of detection and accuracy metrics) of several estimators are performed to understand relative merits and demerits of SFF-based \textit{vs} state-of-the-art GCC-based estimators\footnote{This work is an extension to the preliminary work done in \cite{ncc_2020_sush}}. The tests are conducted on publicly available data collected in real rooms in challenging conditions such as high reverberation and multiple speakers. For rigorous evaluations, the data is further corrupted by different types of noise. From the experiments, it is observed that the detection and accuracy capabilities of the proposed and an existing SFF-based estimator are superior or comparable to those of the best performing GCC-based estimator in all the test cases. \C{Lastly, the advantage of using only the speech dominant bins for robust DoA estimation is demonstrated by testing a \textit{weighted} GCC-PHAT on the corrupted data. Future research would focus on incorporating such weighting functions in the SFF-based methods.}             

\end{abstract}




%

\end{frontmatter}


\section{\label{sec:intro}Introduction}
 Speech processing devices such as hands-free communication kits, voice-controlled, smart-home devices and hearing aids require high quality speech from the target speaker for robust performance. In a closed room environment, the target speech collected at the array of microphones within the devices may be corrupted by background noise, room reverberation and/or interfering speakers. One effective way of enhancing the target speech is to use the array of microphones to, first, find the Time Difference of Arrival (TDoA)/DoA of the signals at the microphones, and then, to use this information of TDoA/DoA to extract the target speech  as in beamformers \cite{BF1988} or as in DoA-informed source extraction techniques \cite{2017_DOAinformed_Maja, 2019_Wood_RT_GCC_NMF, WoodTran,2013_PERTILA_TFsparsityforLoc}. In this paper, the terms DoA and TDoA are used interchangeably because knowing one term, the other one can be estimated from the knowledge of the microphone array geometry (as described in Section ~\ref{sec:signal_model_sff}). \par\indent TDoA estimators may be broadly classified based on the kind of features of speech production used in them as: (i) spectral feature based or (ii) excitation source feature based estimators. The widely used GCC-based estimators \cite{1976_GCC_knapp} exploit spectral features that mostly correspond to the shape of vocal tract during speech production \cite{2005_Hilbert_yegna}. While in GCC, a TDoA is obtained for each time frame in the STFT representation, in a few other estimators, a TDoA is obtained locally in each time-frequency (TF) bin \cite{2004_nbsrp_hist1, 2008_nbsrp_hist2, 2006_mask_Mouba, 2003_nb_hist3_on_spatial_cues,madhu08mog}. While the former estimators are termed as broadband (since all the frequencies per time frame are considered), the latter are called narrowband. \\ 
\indent On the other hand, a few TDoA estimators exploit features from excitation source of speech production. An example of this feature is impulse-like excitations corresponding to epoch locations in a speech signal. While the spectral features are affected by degradations such as noise and reverberation, the excitation features are less affected \cite{2005_Hilbert_yegna}. Therefore, the estimators that exploit excitation source features may be more robust than those that use spectral features. In \cite{2005_Hilbert_yegna}, excitation source features extracted using Hilbert envelope of Linear Prediction (LP) residual of the microphone signal are used for TDoA estimation. The TDoA estimator proposed in \cite{2005_Hilbert_yegna} has \textit{lower or equal} bias, variance and root mean square error than GCC-based estimators, thereby highlighting the benefit of using excitation features.\par\indent Recently, excitation source features emphasized in SFF representation are explored for TDoA/DoA estimation in \cite{2019_sff_loc_sudarsana, ncc_2020_sush}. By filtering a differenced speech signal through a single pole filter, complex outputs (consisting of amplitude envelopes and the corresponding phase) at any desired frequency can be obtained. The pole is placed close to the unit circle in the z-plane to get a very narrow band filter, effectively filtering a single frequency and hence the term ``single frequency filtering (SFF)" is used. The SFF outputs at desired frequencies are stacked to obtain the SFF TF representation. In the SFF domain, speech signals are represented with high temporal resolution of impulse-like events and spectral resolution of harmonics and resonances. This combination of high spectral and temporal resolution is not evident in the STFT representation as demonstrated in Section~\ref{sec:STFTVsSFF}. The high signal-to-noise ratio regions (pertaining to impulse-like excitations) in the mean and variance of SFF outputs across the frequencies are used for DoA estimation in \cite{ncc_2020_sush}. In \cite{2019_sff_loc_sudarsana}, SFF amplitude envelopes at several frequencies are cross-correlated with the corresponding amplitude envelopes at the other channel for robust TDoA estimation. In this SFF-based TDoA estimator, the high signal-to-noise ratio regions at various frequencies of SFF amplitude envelopes of mixture signals are exploited.\C{In the SFF amplitude envelopes, information of impulse-like excitation is embedded with good resolution.} Not only for TDoA estimation but also in other speech applications such as - extraction of fundamental frequency \cite{2017_fundamental_freq_aneeja, 2016_pannala_fo_sff} and epoch extraction \cite{2017_epoch_extraction_kadiri} - the presence of impulse property in amplitude envelopes is exploited. Further, the spectral features emphasized in SFF representation are exploited for TDoA estimation in studies in \cite{interspeech2020_sffphase}, where instantaneous time delays are estimated by using the phase of SFF outputs.\\ 
\indent Based on the facts mentioned above, we hypothesize that TDoA estimators that use excitation source features of speech production would be superior than those that use system-based features of speech production and that SFF TF domain may be better than the STFT TF domain. Therefore, we further experimented with an SFF-based estimator that uses excitation source features \cite{2019_sff_loc_sudarsana} and suggested a modification to it to enhance its performance. Specifically, we propose to perform a PHAT weighted correlation than the time domain correlation performed in \cite{2019_sff_loc_sudarsana}, i.e., SFF amplitude envelopes at various frequencies are cross-correlated with PHAT weighting since, PHAT weighting leads to robust DoA estimators. Lower frequency components of speech, which have significantly higher energy than higher frequency components \cite{comp2014}, would dominate the summation in the GCC-based function, making the peaks in GCC-based function broad and less sharp \cite{2020_thesis_ZhongQiu_Wang}. PHAT weighting flattens the magnitude spectrum, resulting in narrow peaks. The study in \cite{2008WhyPhatWorksInNoise} proves that PHAT works well in reverberation and low noise conditions. \par\indent Moreover, the existing SFF-based estimators in \cite{interspeech2020_sffphase, 2019_sff_loc_sudarsana, ncc_2020_sush} are not evaluated rigorously. For example, only two of the three SFF-based DoA estimators are compared with GCC-based DoA estimators \cite{2019_sff_loc_sudarsana, ncc_2020_sush}. Also, the experiments in \cite{2019_sff_loc_sudarsana} are evaluated on some lab recordings, which lack the diversity of room acoustic conditions. The evaluations in \cite{ncc_2020_sush} are conducted both on simulated and real recordings. But the real data chosen does not have adverse conditions. The metrics chosen to evaluate the three estimators are not exhaustive. To fill these gaps, the present study evaluates the SFF-based and GCC-based estimators more rigorously in terms of several source detection and accuracy metrics on publicly available databases consisting of real recordings. The recordings are further corrupted with different kinds of noise for more rigorous evaluation. Also, unlike the studies in  \cite{interspeech2020_sffphase, 2019_sff_loc_sudarsana, ncc_2020_sush}, a systematic method of choosing various SFF parameters is adapted. These evaluations will help in understanding of relative merits and demerits of various SFF-based compared to GCC-based estimators. Hence, we expect that the results in this paper are more indicative of the comparative performance of SFF-based and GCC-based estimators. Lastly, the performance of a $\textit{weighted}$ GCC-PHAT is compared with the best performing SFF-based methods. The $\textit{weighted}$ GCC-PHAT is obtained by highlighting a few \textit{reliable} TF bins in which the speech sources are dominant. To summarize, the main contributions of the study are:

\begin{itemize} 

\item A method to improve the performance of an existing single frequency filtering (SFF)-based DoA estimator is proposed. On the DoA estimation experiments conducted on publicly available LOCATA and SiSEC datasets, the detection and accuracy capabilities of the DoA estimator with the proposed modification are significantly better.
\item None of the studies compare the existing four SFF-based with the GCC-based estimators rigorously. In this contribution, rigorous (by adding different noises to the SiSEC data) evaluations are performed. Further, a systematic method to choose the parameters of the estimators is developed. 
\item On the rigorous tests conducted, the performance of the proposed (SFF-PHAT-env) and another SFF-based (SFF-PHAT) estimators are superior or comparable to the best-performing GCC-based (GCC-PHAT) estimator in all the metrics. 
\item  The performance of GCC-PHAT is improved by highlighting the speech dominant TF bins and this $\textit{weighted}$ estimator outperforms the \textit{unweighted} GCC-PHAT in all the performance measures on the noisy SiSEC. However, SFF-PHAT has lower MAE metrics than the \textit{weighted} GCC-PHAT and has a comparable F-measure in most of the noises, showing the benefit of SFF-PHAT.
\end{itemize} 

The remainder of the paper is organized as follows: Section~\ref{sec:signal_model_sff} gives the signal model and the relation between TDoA and DoA. Section~\ref{sec:2} describes the method to obtain the SFF  representation. The two TF representations, SFF and STFT, are compared in Section~\ref{sec:STFTVsSFF}. Section~\ref{sec:SFFBasedApproaches} presents a new SFF-based DoA estimator along with the description and implementation details of the existing SFF-based DoA/TDoA estimators. The details of the experimental set-up are given in Section~\ref{sec:Experiments and Results}, which includes the data set, baseline estimators (GCC-based and Hilbert envelope-based) chosen for comparison, along with the parameters chosen, evaluation metrics, and experiments.  The results and discussion are in Section~\ref{sec:ResultsAndDiscussion}. In addition, a $\textit{weighted}$ GCC-PHAT is compared with SFF-PHAT. Section~\ref{sec:conclusions} presents the summary of the study, conclusions, and scope for future work.             

\section{\label{sec:signal_model_sff}Signal Model}Consider an array of \textit{M} microphones that captures the signals radiated by \textit{Q} broadband sound sources in the far field (i.e., the distance of the source from the array is much larger than the array's aperture size). The microphone locations may be expressed in 3D Cartesian co-ordinates by the vectors as $\mathbf{r}_{1}$,\,\ldots,$\mathbf{r}_M$, where $\mathbf{r}_m = [r_{m,x}, r_{m,y}, r_{m,z}]^T$. Under the far field assumption, the DoA vector for source $q$ in this co-ordinate system can be denoted as:
\begin{equation}
    \mathbf{n}_{q}(\theta,\phi) = \begin{pmatrix}\cos(\theta_q)\sin(\phi_q)\,, &  \sin(\theta_q)\sin(\phi_q)\,, & \cos(\phi_q)\end{pmatrix}^T\,,
\end{equation}
where $0 \leq\theta\leq2\pi$ is the azimuth angle between the projection of $\mathbf{n}_{q}(\theta,\phi)$ on to the $xy$ plane and the positive $x$-axis and $0\leq\phi\leq\pi$ is the elevation angle with respect to the positive $z$-axis.\\ 
In the STFT domain, the image of source $q$ at the array, in the $k^{\text{th}}$ frequency bin and $b^{\text{th}}$ time frame, can be compactly denoted as: 
$\mathbf{X}_q(k,b)=[X_{q,1}(k,b),\, \ldots,X_{q,M}(k,b)]^T\,.$
If $\mathbf{V}(k,b)$ is the STFT-domain representation of the background noise at the array, the net signal captured by the array can be written as:
\begin{equation}\label{eq:atf}
    \mathbf{X}(k,b) = \sum_{q=1}^{Q}\mathbf{X}_{q}(k,b) +   \mathbf{V}(k,b),
\end{equation} 
where $\mathbf{X}(k,b)=[X_{1}(k,b),\, \ldots,X_{M}(k,b)]^T\,.$\\ 
\subsection{\label{subsec:tdoa}Relation between DoA and TDoA in the Case of Two Microphones}
\C{In linear arrays, DoA estimation  refers to the task of finding the broadside angle, which is the angle between the plane orthogonal to the linear array axis and the source DoA vector. The broadside angle is equivalent to the azimuth angle $\theta$ based on the following reasoning: the point source and the linear array are co-planar because any point (i.e., source location) in three-dimensional space can be viewed as co-planar with a line (i.e., array axis). Hence, finding the broadside angle is equivalent to finding only the azimuth angle $\theta$ on the plane containing the source and the array. The elevation angle $\phi$ is, then, $90^o$. To summarize, in linear arrays, DoA refers to the azimuth angle $\theta$, and in planar arrays, it refers to both $\theta$ and $\phi$. \C{Without loss of generality, in the works in the thesis, we consider localization in the azimuth plane only (i.e., $\phi=\pi/2$).}}

In two microphone DoA estimators, the usual first step is to estimate the time difference of arrival of the signals at two microphones. Subsequently, DoA of the signal of interest is estimated from the TDoA $\tau$ and with the knowledge of microphone array geometry as follows: the microphone positions in 3D Cartesian co-ordinates can be expressed in vector notation as $\mathbf{r}_1$ and $\mathbf{r}_2$. In case of two microphone, $\phi = \frac{\pi}{2}$ because it may be assumed that the sound source and the linear array are co-planar and the DoA corresponds to $\theta$. Then, DoA vector  $\mathbf{n}_{q}(\theta,\frac{\pi}{2}) = \begin{pmatrix}\cos(\theta_q)\sin(\frac{\pi}{2})\,, &  \sin(\theta_q)\sin(\frac{\pi}{2})\,, & \cos(\frac{\pi}{2})\end{pmatrix}^T\ = \begin{pmatrix}\cos(\theta_q)\,, &  \sin(\theta_q)\,, & 0\end{pmatrix}^T\,$. DoA vector $\mathbf{n}_{q}(\theta,\frac{\pi}{2})$ and TDoA $\tau$ are related as follows:
\begin{equation}
    \tau = \frac{[\mathbf{r}_1 - \mathbf{r}_2]^T\mathbf{n}_{q}(\theta,\frac{\pi}{2})}{c},
    \label{eq:doatau}
\end{equation} 
where, c is speed of sound at room temperature.
Given the array geometry and the estimated $\tau$, using Equation ~\eqref{eq:doatau}, DoA $\theta$ can be estimated. 

\C{The estimators in \cite{interspeech2020_sffphase, 2019_sff_loc_sudarsana} are proposed to estimate the time difference of arrival (TDoA) between the target signal at two microphones rather than to estimate DoA of the target signal. However, from a given TDoA, corresponding DoA may be obtained if the microphone array geometry is known.}  
Since the DoA can be calculated from the estimated TDoA, the two terms are used interchangeably. Henceforth, all the estimators would be referred to as DoA estimators, though they may have been proposed for TDoA estimation. All the correlation functions in the following sections would be defined as a function of azimuth angle $\tau(\theta)$, where $\theta$ is set of pre-selected search DoAs over which the estimator is swept through for DoA estimation.

\section{\label{sec:2}SFF TF representation of Speech Signals}
By passing the frequency-shifted speech signals through a near ideal resonator located at half the sampling frequency $\frac{f_s}{2}$, the complex SFF output (envelope and the corresponding phase) as a function of time is obtained at any desired frequency. The steps involved in computing the SFF output at a given frequency $f_k$ are as follows  \cite{sff}:
\begin{enumerate}

\item The microphone speech signal $x[n]$ is differenced to reduce low frequency trend in the signal as follows:
\begin{equation}
\tilde{s}[n] = x[n] - x[n-1].
\end{equation}
\item The differenced signal $ \tilde{s}[n]$ is frequency-shifted by multiplying it with $e^{j\bar{\omega}_k{n}}$
(where $\bar{\omega}_k = \pi-{\omega}_k=\pi-\frac{2 \pi f_k }{f_s}$).
The frequency-shifted signal is given by:
\begin{equation}
 \tilde{s}[k,n]=\tilde{s}[n]e^{j\bar{\omega}_k{n}}.
\end{equation}
\item The signal $\tilde{s}[k,n]$ is passed through a single pole filter with transfer function given by:

\begin{equation}
\label{eq:SFF_zplane}
 F_{\text{SFF}}(z)= \frac{1}{1+rz^{-1}}.
\end{equation} 
The root is on the negative real axis at $z = -r$. To ensure the stability of the filter, the $r$ value is chosen closer to the unit circle ($r \approx 1$). This corresponds to filtering the signal using a near ideal resonator at $f_s/2$.

\item The output of the filter is given by:
\begin{equation}
  y[k,n] = -ry[k,n-1]+\tilde{s}[k,n].
\end{equation} 

\item Let  $y[k,n] = y_{r}[k,n] + jy_{i}[k,n]$, where  $y_{r}[k,n]$ and $y_{i}[k,n]$ are the real and imaginary parts of $y[k,n]$. The magnitude or amplitude envelope $e[k,n]$ and the phase $\psi[k,n]$ of the signal $y[k,n]$ at $k^{th}$ frequency are given by:
\begin{equation}
 ~~~~~~~~~~ e[k,n]=\sqrt{y_{r}^2[k,n] + y_{i}^2[k,n]}, 
\end{equation} 
and 
\begin{equation}
~~~~~~~~~~ \psi[k,n]=tan^{-1}(\frac{y_{i}[k,n]}{y_{r}[k,n]}).
\end{equation} 

\end{enumerate}

The impulse response of the filter in Equation~\eqref{eq:SFF_zplane} is an exponentially decaying window $f_{SFF}[n]=(-r)^{n}u[n]$, where $u[n]$ is a unit step function. As the value of $r$ increases, the 3 dB bandwidth of the filter ${B_{SFF}}$ decreases. The parameter $r$ and the $3~dB$ bandwidth of the filter ${B_{SFF}}$ (in $Hz$) are inversely related by ~\cite{2019_icassp_krishna}:
\begin{equation}
    {B_{SFF}} = \frac{f_s}{2\pi}cos^{-1}\big(\frac{4r-r^2-1}{2r}\big).
\end{equation}         
The SFF output is obtained for $K_{SFF}$  frequencies that are equally spaced intervals between $0$ and $\frac{f_s}{2}$.  
From the amplitude envelopes of SFF $e[k,n]$, we can get the SFF spectrum of the signal at every instant of time.   
The important characteristic of $e[k,n]$ at different frequencies $f_k$ is that it has some high  SNR regions (impulse-like events corresponding to epochs) \cite{sff}. \C{This is due to correlation among the speech samples and lack of correlation among the noise samples.} This high SNR property is exploited in DoA estimators. 
   
\section{\label{sec:STFTVsSFF}Comparison of SFF and STFT TF representations}

\begin{figure}[h!]
 \centering
    \includegraphics[width=0.9\columnwidth] {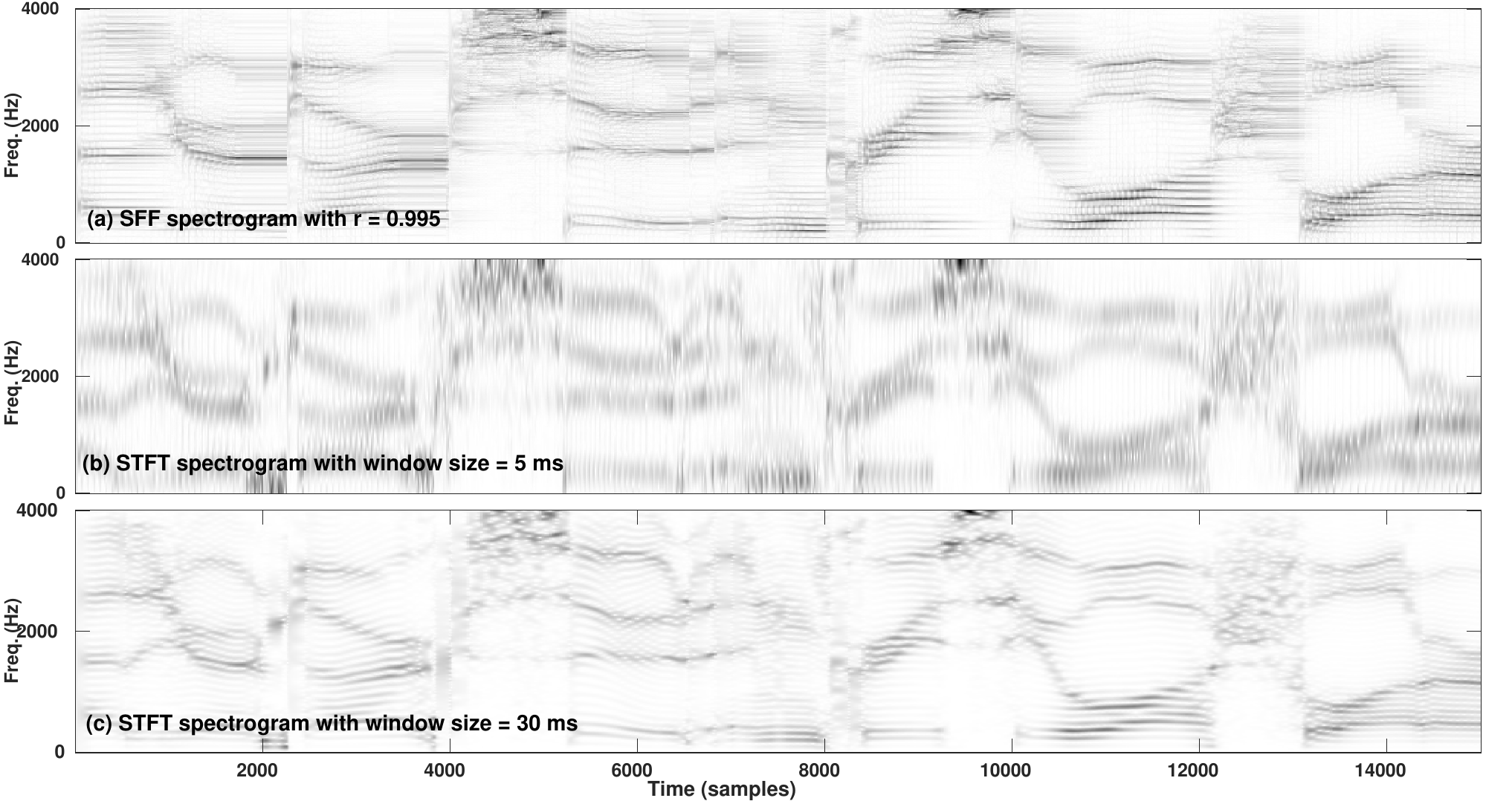}
    \caption[Demonstration of Benefit of SFF over STFT - Figure 1]{(a) SFF spectrogram obtained with $r=0.995$ and $
    K_{SFF}=200$ and the speech signal is sampled at 8 KHz. (b) STFT spectrogram obtained with a Hann window of 5~ms with a hop of 1 sample and 400 point DFT. (c) STFT spectrogram obtained with a Hann window of 30~ms with a hop of 1 sample and 400 point DFT. It is observed that the spectral resolution in (a) and (c) is significantly better than the spectral resolution in (b).}  
    \label{fig:SFF and STFT comparison1}
\end{figure}

\begin{figure}[h!]
 \centering
    \includegraphics[width=0.9\columnwidth] {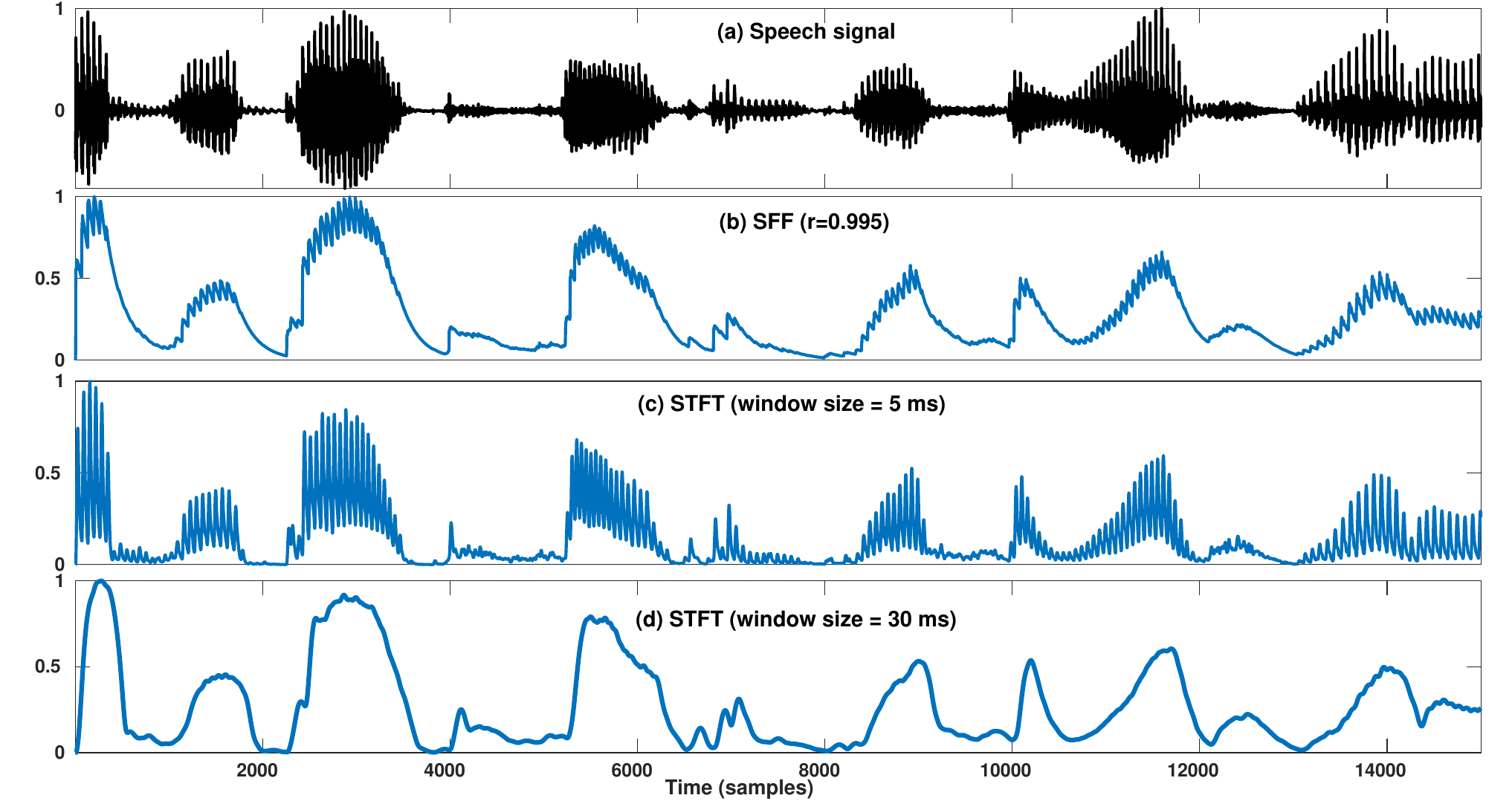}
    \caption[Demonstration of Benefit of SFF over STFT - Figure 2]{(a) A Speech signal. (b) Mean (across frequencies) of SFF spectrogram of the signal in (a). (c) Mean (across frequencies) of STFT spectrogram of the signal in (a) obtained with a short window of 5~ms. (d)  Mean (across frequencies) of STFT spectrogram of the signal in (a) obtained with a long window of 30~ms. The vertical striations in (b) and (c) indicate good temporal resolution, implying that the STFT with short window and SFF capture the occurrence of impulse-like excitations. The absence of vertical striations in (d) indicate poor temporal resolution of long window STFT.}  
    \label{fig:SFF and STFT comparison2}
\end{figure}

\begin{figure}[h!]
 \centering
    \includegraphics [width=0.9\columnwidth]{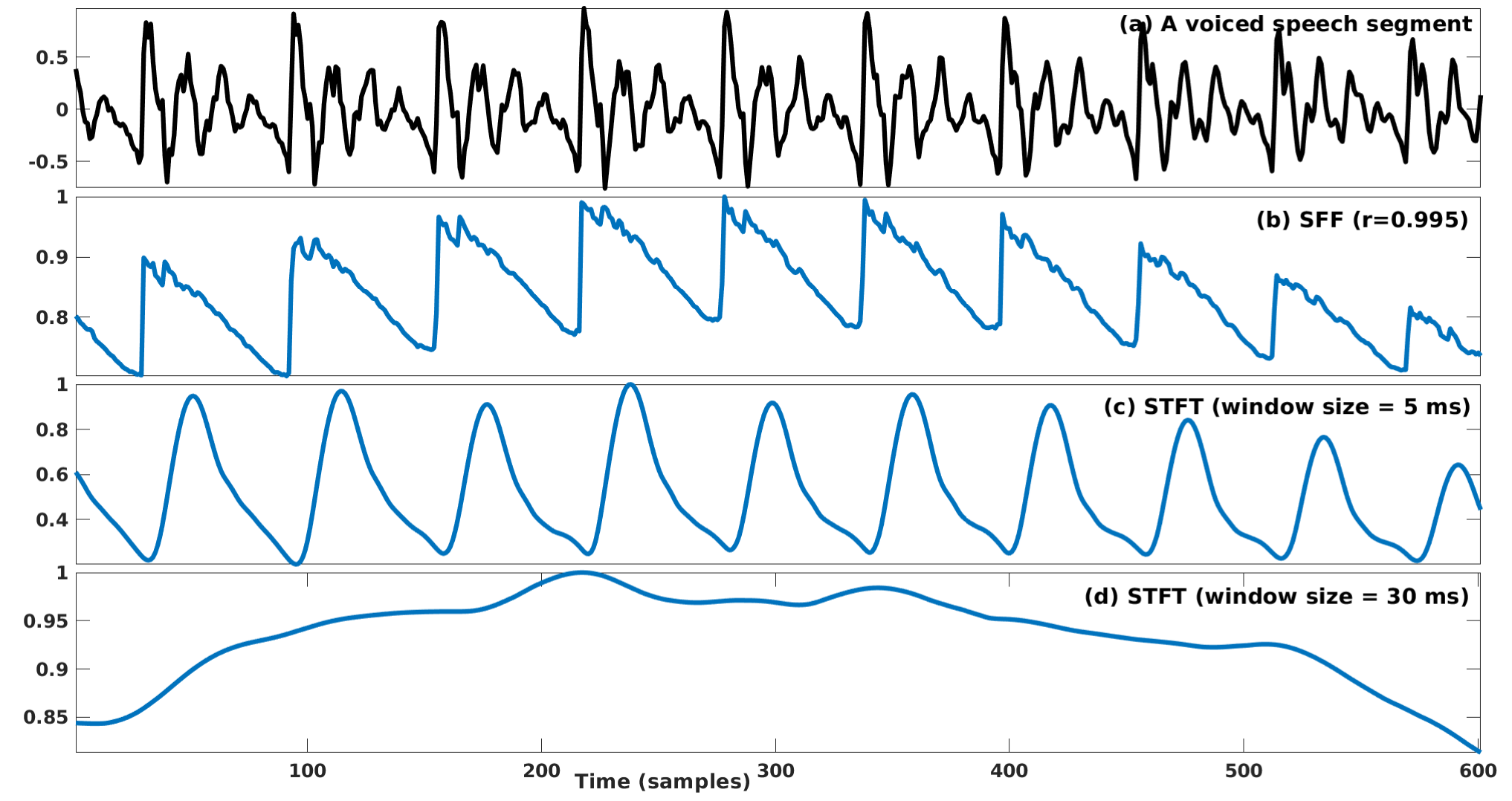}
    \caption[Demonstration of Benefit of SFF over STFT - Figure 3]{(a) A voiced speech segment (Slice of the speech signal in Figure~\ref{fig:SFF and STFT comparison2}). (b) Mean (across frequencies) of SFF spectrogram of the segment in (a). (c) Mean (across frequencies) of STFT spectrogram of the segment in (a) obtained with a short window of 5~ms. (d)  Mean (across frequencies) of STFT spectrogram of the segment in (a) obtained with a long window of 30~ms. SFF (b) and STFT with short window (c) capture occurrence of impulse-like excitations, implying good temporal resolution. STFT with long window (d) has poor temporal resolution.}  
    \label{fig:SFF and STFT comparison3}
\end{figure}

SFF representation of speech provides high temporal resolution of some features of excitation source such as impulse-like events, and high spectral resolution of features such as harmonics \cite{2017_epoch_extraction_kadiri, 2016_pannala_fo_sff}. The high spectral and temporal resolution of SFF over STFT may be visualized using the Figures~\ref{fig:SFF and STFT comparison1} to~\ref{fig:SFF and STFT comparison3}.  

\clearpage

Figure~\ref{fig:SFF and STFT comparison1} shows the spectrograms of SFF and STFT of a speech signal. SFF spectrogram in Figure~\ref{fig:SFF and STFT comparison1}(a) is obtained with $r = 0.995$ and $K_{SFF} = 200$. STFT spectrogram is obtained with a 400-point DFT so that the unique frequency bins of STFT ($\approx 200$) is equal to $K_{SFF}$. Hann window of sizes 5~ms in Figure~\ref{fig:SFF and STFT comparison1}(b) and 30~ms in Figure~\ref{fig:SFF and STFT comparison1}(c) at sampling rate of 8 kHz and hop size of $1$ sample are used. From the figure, it is observed that the spectral resolution of SFF and that of STFT with 30~ms is higher than the spectral resolution of STFT with 5~ms. 
To visualize the temporal resolution, the spectrograms are summed across frequencies and plotted in Figure~\ref{fig:SFF and STFT comparison2} along with the speech signal. In Figure~\ref{fig:SFF and STFT comparison3}, a portion of Figure~\ref{fig:SFF and STFT comparison2} is selected and plotted. 

From Figures~\ref{fig:SFF and STFT comparison2} and~\ref{fig:SFF and STFT comparison3}, it is evident that the temporal resolution of impulse-like events is high in SFF and STFT with window size of 5~ms, and in STFT with window size of 30~ms the temporal resolution is poor. Hence, it can be concluded that SFF offers a high spectral and temporal resolution which is not possible in STFT with a single window size. \C{The high resolution of SFF has been exploited for robust  fundamental frequency estimation \cite{2016_pannala_fo_sff, 2017_fundamental_freq_aneeja}, epoch extraction \cite{2017_epoch_extraction_kadiri} and DoA/TDoA estimation \cite{interspeech2020_sffphase, 2019_sff_loc_sudarsana, ncc_2020_sush}. Motivated with this, the present study investigates and compares the performance of DoA/TDoA estimators in STFT and SFF representations.}

\section{\label{sec:SFFBasedApproaches} SFF-based DoA Estimators}
In SFF TF representation, the DoA estimators exploit either spectral features \cite{interspeech2020_sffphase} or excitation source features \cite{2019_sff_loc_sudarsana, ncc_2020_sush}. 
Cross-correlation of mean and variance spectral features derived from SFF representation is performed in \cite{ncc_2020_sush}, where, the robust high SNR regions in the spectral features are exploited for accurate DoA estimation.These estimators are referred to as SFF-mean and SFF-var respectively. The implementation details of SFF-mean and SFF-var are in Section ~\ref{subsec:descriptionOfSFFmeanSFFvar}. SFF amplitude envelopes obtained at several frequencies are cross-correlated across the channels for TDoA estimation in \cite{2019_sff_loc_sudarsana}, in which the high SNR regions in the SFF amplitude envelopes and multiple evidences at several frequencies are jointly utilized for robust TDoA estimation. The details of implementation of this estimator, called SFF-env, are given in Section ~\ref{subsec:descriptionCSSP}. In \cite{interspeech2020_sffphase}, the phase of SFF outputs is utilized for robust TDoA estimation just as STFT phase is used in GCC-PHAT. The estimator in \cite{interspeech2020_sffphase} is referred as SFF-PHAT and is described in Section ~\ref{subsec:descriptionSffphase}.\par\indent Between spectral and excitation source features, the latter features are robust to noise and reverberation \cite{2005_Hilbert_yegna}. Therefore, we propose a modification to DoA estimator in \cite{2019_sff_loc_sudarsana} in which the excitation source features are used. The new SFF-based DoA estimator is described in Section ~\ref{subsec:SFF-PHAT-env} and is referred as SFF-PHAT-env, in which SFF amplitude envelopes at various frequencies are cross-correlated with PHAT weighting. In this estimator, the high SNR regions at SFF amplitude envelopes, multiple evidences at several frequencies and robust PHAT weighting are exploited for accurate DoA estimation.  \\


\subsection{\label{subsec:SFF-PHAT-env}Proposed SFF-Based DoA Estimator: SFF-PHAT-env}
\begin{figure}[h!]
  \centering
  \includegraphics[width=1\columnwidth]{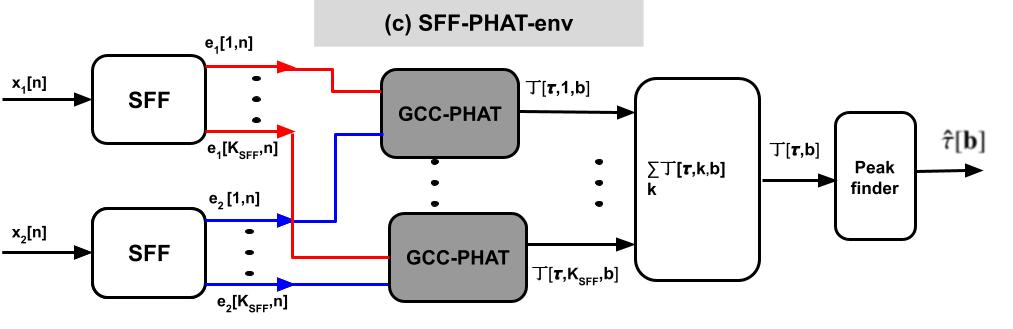}
  \caption[Block diagram of the proposed SFF-based DoA estimator: SFF-PHAT-env]{Block diagram of the proposed SFF-based DoA estimator: SFF-PHAT-env. The grey block is the modified block when compared to the SFF-env implementation. The GCC-PHAT operation is presented in Equation~\eqref{eq:bb_gcc_phat}; the objective function is defined in Equation~\eqref{eq:cc_sff}; and the DoA can be estimated from the TDoA as described in Section~\ref{sec:signal_model_sff}.}  
  \label{fig:sff_PHAT_env}
\end{figure}

The key idea in SFF-PHAT-env estimator is to cross correlate SFF outputs of the microphone signals at several corresponding frequencies using PHAT weighting. The advantage of the SFF output is that it has high signal-to-noise regions in both time and frequency domains. Further, this estimator gives multiple evidences, one from each of the SFF output frequencies. These multiple evidences are combined to obtain robust DoA estimates. Additionally, the PHAT weighting makes it robust in adverse conditions. 
Figure~\ref{fig:sff_PHAT_env} shows the block diagram of SFF-PHAT-env implementation. 

\par 
Let $e_{1}[k,n]$ and $e_{2}[k,n]$ be the SFF amplitude envelopes obtained at the $k^{th}$ frequency of two microphone signals $x_{1}[k,n]$ and $x_{2}[k,n]$ respectively. At each frequency $k$, the amplitude envelopes at the two channels are correlated with PHAT weighting, resulting in a cross correlation function as follows: 
\begin{equation}
 \mathcal{J}_{\text{SFF-PHAT-env}}[\tau(\theta),k,b] = \text{GCC-PHAT}(e_1[k,n],e_2[k,n]),
 \label{eq:cc_sff}
\end{equation}
where GCC-PHAT is defined in Equation~\eqref{eq:bb_gcc_phat}.
$\mathcal{J}_{\text{SFF-PHAT-env}}[\tau(\theta),k,b]$ at each frame are summed across frequencies as:
\begin{equation}
\mathcal{J}_{\text{SFF-PHAT-env}}[\tau(\theta),b] = \sum\limits_{k=1}^{K_{SFF}}\mathcal{J}_{\text{SFF-PHAT-env}}[\tau(\theta),k,b].
\end{equation}
At each time frame $b$, the peak location in $\mathcal{J}_{\text{SFF-PHAT-env}}[\tau(\theta),b]$ is estimated, indicating the DoA of an active speaker in that time frame (frame-wise DoA) as: 

\begin{equation}
    \hat{\theta}(b) =  {\argmax_{\theta}}~~ \mathcal{J}_{\text{SFF-PHAT-env}}[\tau(\theta),b].
\end{equation}
The frame-specific DoA estimates $\hat{\theta}(b)$ are used to obtain a histogram. The peak locations in the histogram represent the source DoA estimates. It is assumed that the number of speakers is known \textit{a priori}. The locations of a few top peaks (corresponding to the number of speakers) in the histogram indicate the DoA of the sources in the mixture. A graphical illustration of obtaining source DoAs from histogram of frame-wise DoA estimates is shown in Figure~\ref{fig:SiSEC_fem4_250ms_sff_phase}.

\subsection{\label{subsec:descriptionOfSFFmeanSFFvar}SFF-mean and SFF-var}
\begin{figure}[h]
  \centering
  \includegraphics[width=0.9\textwidth,height=8cm]{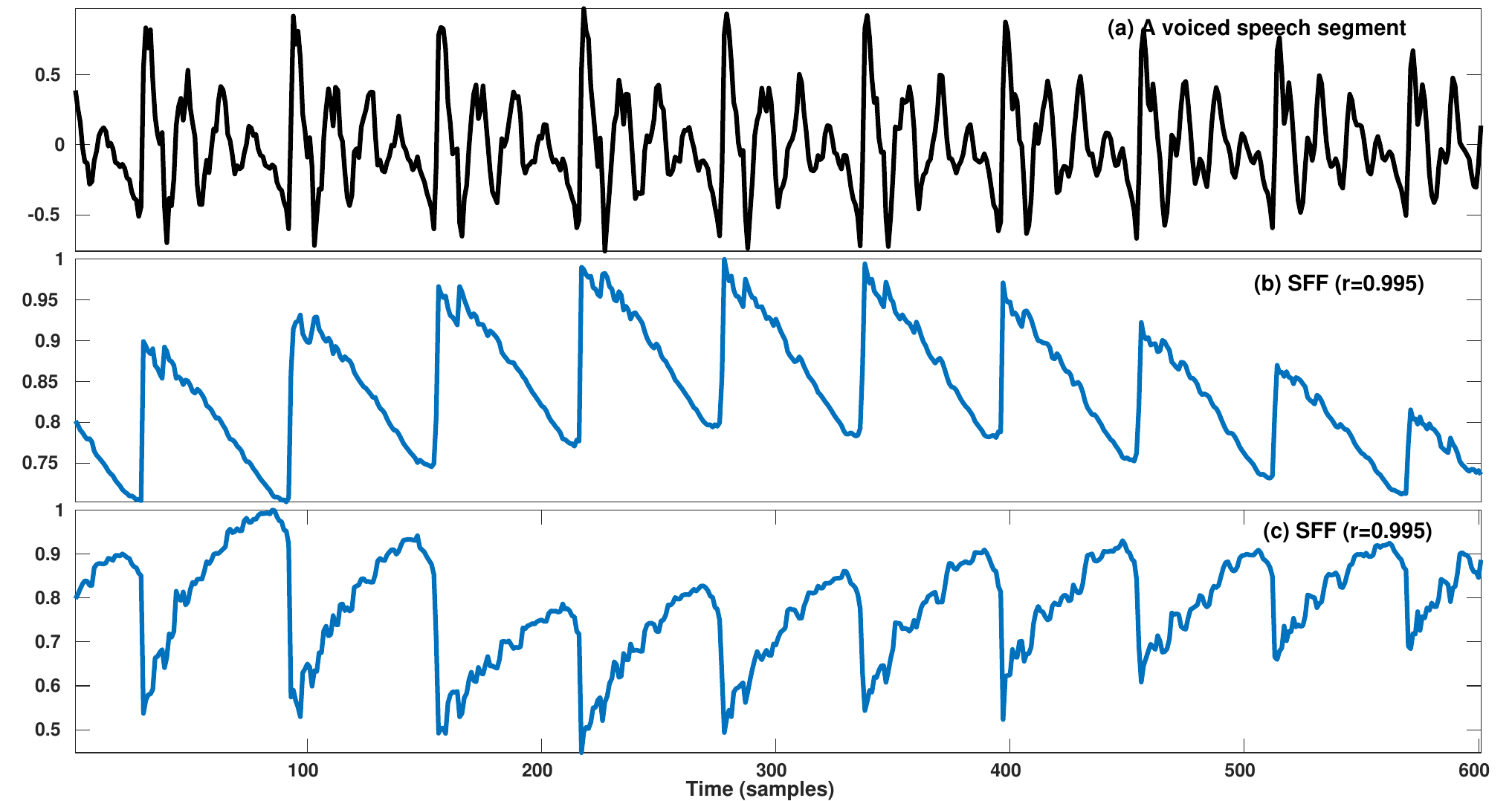}
  \vspace{-0.2cm}
  \caption[Mean and Variance of SFF amplitude envelopes]{(a) A segment of voiced speech. (b) Mean of SFF amplitude envelopes of the voiced segment in (a). (c) The variance of SFF amplitude envelopes of the voiced segment in (a). SFF amplitude envelopes are obtained with $K_{SFF} = 256$ and $a=0.995$. It is observed that the mean and variance spectral features highlight the locations of high SNR impulse-like excitations present in the voiced segment. The peaks in (b) and the valleys in (c) coincide with the occurrence of impulse-like events in (a).} 
  \label{fig:MeanAndVar}
  \vspace{-0.1cm}
\end{figure}

The key idea of these estimators, which are proposed in \cite{ncc_2020_sush}, is to exploit the high SNR regions in the SFF representation for robust DoA estimation. It is observed that spectral features derived from SFF representation highlight the high SNR regions. In particular, the spectral features explored are the mean and variance of the SFF amplitude envelopes across frequencies. It is demonstrated in Figure~\ref{fig:MeanAndVar} that the mean and variance capture the high SNR regions. Figure~\ref{fig:MeanAndVar} (a) shows a segment of voiced speech,~\ref{fig:MeanAndVar} (b) and (c) show the mean and variance (across frequencies) of its SFF amplitude envelopes. From Figure~\ref{fig:MeanAndVar} (b), it is evident that the mean $\mu[n]$  captures high SNR regions - the peaks in spectral mean are indicators of high SNR regions. This is because high energy impulse-like excitations, when averaged across frequencies, lead to peaks. On the other hand, the valleys in spectral variance $\sigma^2[n]$ are indicators of high SNR regions. This is because an impulse-like excitation has a flat spectrum, leading to low values in the variance across frequencies.        

\indent The  mixture  signals collected at the microphones are  transformed  into SFF  representation. Subsequently,  spectral features obtained from the  temporal  amplitude envelopes  across  frequencies are  correlated across the channels to get robust  DoA  estimates  in real acoustic environments. The DoA estimators implemented using spectral mean and variance are called SFF-mean and SFF-var, respectively \cite{ncc_2020_sush}. The implementation details of these DoA estimators are described next. 

\subsubsection{\label{subsubsec:SFF-mean}DoA Estimator from Mean Spectral feature in SFF Representation: SFF-mean} 
\begin{figure}[h!]
  \centering
  \includegraphics[width=1\columnwidth]{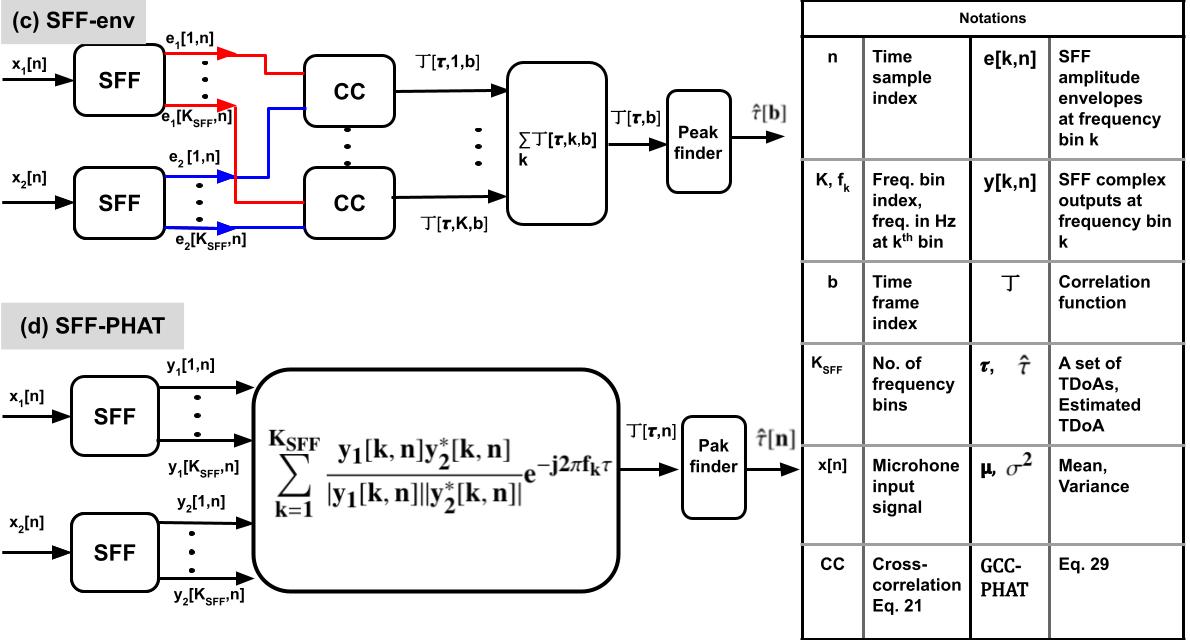}
  \caption[Block diagrams of the existing SFF-based DoA estimators]{Block diagrams of the existing SFF-based DoA estimators: (a) SFF-mean \cite{ncc_2020_sush}, (b) SFF-var \cite{ncc_2020_sush}, (c)SFF-env \cite{2019_sff_loc_sudarsana}, and (d) SFF-PHAT \cite{sff_phase}. $\hat{e}$ and $\hat{\mu}$ are defined in Section~\ref{subsubsec:SFF-var}. $\hat{\theta}[b]$ can be estimated from $\hat{\tau}[b]$ as described in Section~\ref{sec:signal_model_sff}.} 
  \label{fig:all_sff}
\end{figure}

In Figure~\ref{fig:all_sff}(a), a block diagram implementation of SFF-mean is shown. Spectral mean $\mu[n]$ is the mean of the SFF amplitude envelopes and is given by:
\begin{equation}
\label{eq:sffmean}
 \mu[n] = \frac{1}{K_{SFF}}{\sum_{k=1}^{K_{SFF}}{e[k,n]}}.
\end{equation}
The spectral means obtained  at two microphones $\mu_1[n]$ and $\mu_2[n]$ are correlated using GCC-PHAT to obtain the cross correlation function $\mathcal{J}_{\text{SFF-mean}}[\tau(\theta),b]$:

\begin{equation}
   \label{eq:xcorr1_mean}
    \mathcal{J}_{\text{SFF-mean}}[\tau(\theta),b] = \text{GCC-PHAT}(\mu_1[n],\mu_2[n]).   
\end{equation}
At each time frame $b$, the peak location in $\mathcal{J}_{\text{SFF-mean}}[\tau(\theta,b)]$ is estimated, indicating the DoA of an active speaker in that time frame (frame-wise DoA) as: 
 \begin{equation}
   \label{eq:xcorr2_mean}
    \hat{\theta}(b) =  {\argmax_{\theta}}~~ \mathcal{J}_{\text{SFF-mean}}[\tau(\theta),b].
\end{equation}
Histogram of frame-wise DoA estimates $\hat{\theta}(b)$ is obtained. The peak locations in the histogram correspond to the source DoA estimates as described in Section~\ref{subsec:SFF-PHAT-env}.  
\begin{figure}[h!]
  \includegraphics[width=0.8\columnwidth]{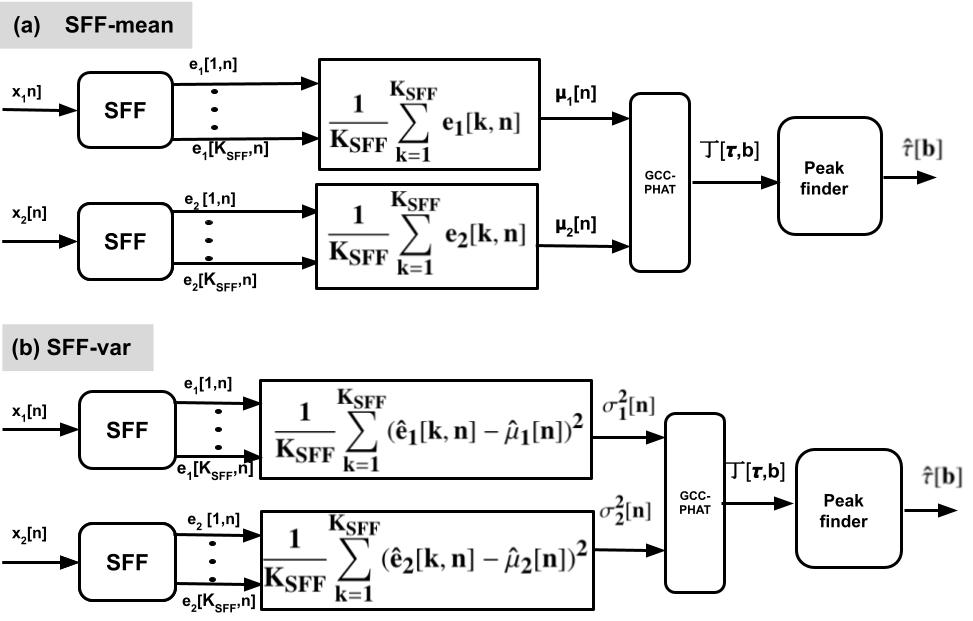}
\end{figure}

\subsubsection{\label{subsubsec:SFF-var}SFF-var}
In Figure~\ref{fig:all_sff}(b), a block diagram implementation of SFF-var is shown. The spectral variance is calculated from normalized amplitude envelope $\hat{e}[k,n]$ and is given by: 
\begin{equation}
\label{eq:sffvar}
 \sigma^2[n] = \frac{1}{K_{SFF}} {\sum\limits_{k=1}^{K_{SFF}}(\hat{e}[k,n]-\hat{\mu}[n])^2},
\end{equation}
where $\hat{e}[k,n] = \frac{e[k,n]}{\sum_{k=1}^{K_{SFF}}e[k,n]}$ and $\hat{\mu}[n]=\frac{1}{K_{SFF}}\sum\limits_{k=1}^{K_{SFF}}\hat{e}[k,n]=\frac{1}{K_{SFF}}$, as the envelopes are normalized across frequency at each time instant.
The spectral variances obtained  at two microphones ($\sigma^2_1[n]$ and $\sigma^2_2[n]$) are correlated using GCC-PHAT to obtain the cross correlation function $\mathcal{J}_{\text{SFF-var}}[\theta,b]$ as: 
\begin{equation}
   \label{eq:xcorr1_var}
    \mathcal{J}_{\text{SFF-var}}[\tau(\theta),b] = \text{GCC-PHAT}(\sigma^2_1[n],\sigma^2_2[n]).  
\end{equation}
At each time frame $b$, the peak location in $\mathcal{J}_{\text{SFF-var}}[\tau(\theta),b]$ is estimated, indicating the time delay of an active speaker in that time frame as: 
 \begin{equation}
   \label{eq:xcorr2_var}
    \hat{\theta}(b) =  {\argmax_{\theta}}~~ \mathcal{J}_{\text{SFF-var}}[\tau(\theta),b].
\end{equation}


The frame-specific DoA estimates $\hat{\theta}(b)$ are used to obtain a histogram.   The peak locations in the histogram correspond to the source DoA estimates as described in Section~\ref{subsec:SFF-PHAT-env}.\par
\begin{figure}[h!]
  \centering
  \includegraphics[width=0.6\columnwidth]{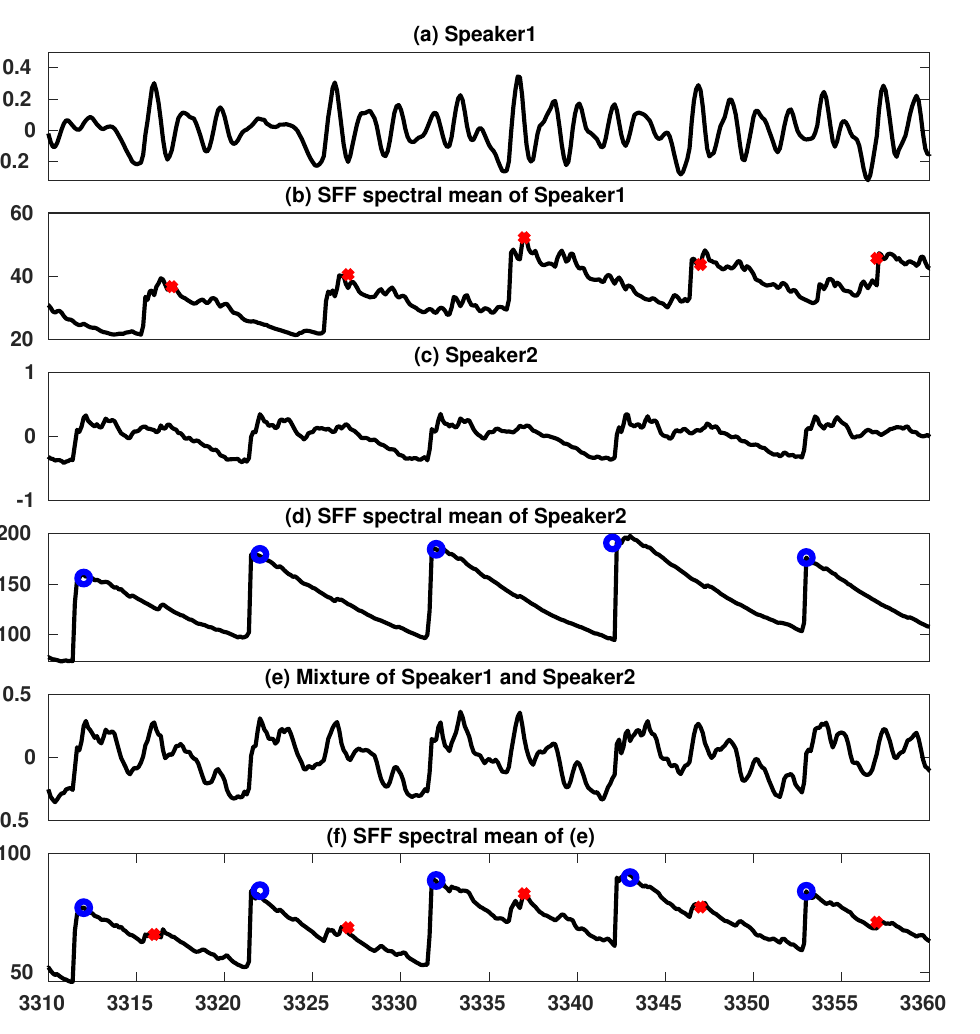}
    \vspace{-0.7cm}

   \center{{\bf{\scriptsize Time (ms)}}}

  \vspace{-0.1cm}
  \caption[Illustration of high SNR spectral features of two concurrent speakers]{(a) and (c) show a voiced segment of Speaker 1 and 2 respectively; (b) and (d) show the mean of SFF amplitude envelopes of Speaker 1 and 2 shown in (a) and (c); (e) shows the segment with both Speaker 1 and Speaker 2  active; and (f) shows the mean of SFF amplitude envelopes derived from (e).  (adapted from \cite{ncc_2020_sush}).}
  \label{fig:MSP and Var singleSpeaker}
  \vspace{-0.1cm}
\end{figure}

\indent In Figure \ref{fig:MSP and Var singleSpeaker}, high SNR spectral features for two concurrent speakers is illustrated.
SFF spectral mean property on voiced segments of two concurrent speakers is shown in the figure. In a simulated room environment, two concurrent speakers' data is captured by a pair of microphones 1m apart. The plots in the figure are obtained from the signals at one of the microphones. Figure \ref{fig:MSP and Var singleSpeaker}(a) shows a voiced segment of a speaker 1. SFF spectral mean of the voiced segment of speaker one is plotted in Figure \ref{fig:MSP and Var singleSpeaker}(b), in which the red markers highlight the time instants of occurrence of impulse-like events of speaker 1. Figure \ref{fig:MSP and Var singleSpeaker}(c) shows  a voice segment of Speaker 2 and Figure \ref{fig:MSP and Var singleSpeaker}(d) shows SFF spectral mean of speaker 2. The instants of impulse-like excitations of speaker 2 are in blue markers. The concurrent speech of speaker one and speaker two is shown in Figure \ref{fig:MSP and Var singleSpeaker}(e). Figure \ref{fig:MSP and Var singleSpeaker}(f) is the SFF spectral mean of the mixture signal in \ref{fig:MSP and Var singleSpeaker}(e). High SNR regions of speakers 1 and 2 are highlighted by red and blue markers, respectively, which correspond to the peaks of the spectral mean. It may be observed from Figure \ref{fig:MSP and Var singleSpeaker}(e) and Figure \ref{fig:MSP and Var singleSpeaker}(f) that while the peaks in Figure \ref{fig:MSP and Var singleSpeaker}(f) are prominent, such prominent peaks are absent in Figure \ref{fig:MSP and Var singleSpeaker}(e). Further, the red and blue markers shown in Figure \ref{fig:MSP and Var singleSpeaker}(f) are aligned across the plots in Figure \ref{fig:MSP and Var singleSpeaker}(b) and Figure \ref{fig:MSP and Var singleSpeaker}(d), respectively, implying that the impulse-like events of the two speakers in  \ref{fig:MSP and Var singleSpeaker}(f) closely match those of individual speakers. Since spectral means in \ref{fig:MSP and Var singleSpeaker}(f)) highlight the impulse-like events (where the effects of noise and reverberation are reduced), correlation of spectral means across the channels would result in more prominent peaks in the cross correlation function than the correlation of mixture signals (Figure \ref{fig:MSP and Var singleSpeaker}(e)). Cross correlation of SFF spectral variance would also result in prominent peaks, but instead of peaks in SFF mean, the valleys in spectral variance correspond to impulse-like events.\\

\subsection{\label{subsec:descriptionCSSP}SFF-env}
\begin{figure}[h]
\centering
\includegraphics[width=0.9\textwidth,height=10cm]{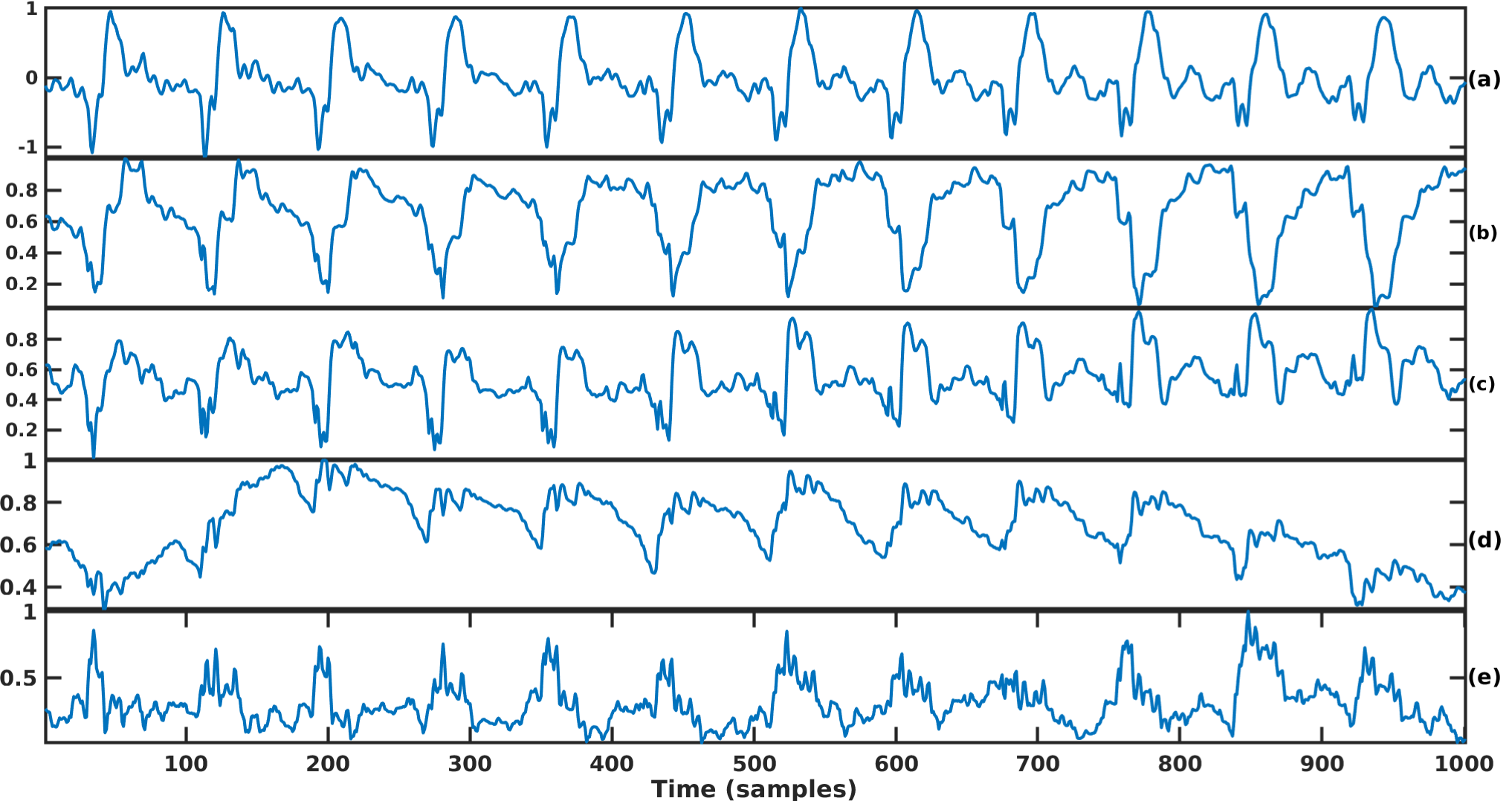}

\caption[Illustration of SFF amplitude envelopes at various frequencies.]{(a) A segment of voiced speech. (b) to (e) shows SFF amplitude envelopes computed at 500 Hz, 1000 Hz, 2000 Hz, and 3000 Hz, respectively. It is observed that the information of high SNR regions is present at various frequencies.}
\label{fig:frequencies}

\end{figure}

The key idea in SFF-env estimator, which is proposed in \cite{2019_sff_loc_sudarsana}, is to cross correlate corresponding SFF outputs of the microphone signals at several frequencies, rather than using the waveforms directly. The advantage of the SFF output is that it will have high signal-to-noise regions in both time and frequency domains. Also, this estimator gives multiple evidences, one from each of the SFF outputs. These multiple evidences are combined to obtain robust DoA estimates. 
Figure~\ref{fig:frequencies} illustrates that the SFF amplitude envelopes at various frequencies have emphasized high SNR regions. The figure shows a segment of voiced speech, and SFF envelopes at 500 Hz, 1000 Hz, 2000 Hz, and 3000 Hz. It is observed that amplitude envelopes at various frequencies highlight the impulse-like locations.   
\par 
In Figure~\ref{fig:all_sff}(c), a block diagram implementation of SFF-env is shown. Typically, cross correlation function at each frequency $k$ is estimated on overlapping frames of amplitude envelopes with frame length between 50~ms to 500~ms so that each frame contains at least a few pitch periods to obtain a good time delay estimate \cite{2005_Hilbert_yegna}. Smaller frame lengths lead to better tracking, while the larger frame lengths give more accurate estimates. Let $e_{1b}[k,n]$ and $e_{2b}[k,n]$ be two corresponding frames each of length $N$ samples at $k^{th}$ frequency bin. The cross correlation function of these frames is given by: 

\begin{equation}
	\label{eqn:cssp}
	\mathcal{J}_{\text{SFF-env}}[\tau(\theta),k,b] =\sum\limits_{n=0}^{\mathcal{N}-1}e_{1b}[k,n]e_{2b}[k,n+\tau(\theta)], \qquad 
	-T_{max}\leq \tau(\theta) \leq T_{max},
\end{equation}
where $T_{max}$ is the maximum time-delay between the given pair of microphones. The cross correlation function is estimated at each frame $b$ where $0 \leq b \leq B$. Each block begins at $b\Delta T$, where $\Delta T$ is the hop size in samples.
$\mathcal{J}_{\text{SFF-env}}[\tau(\theta),k,b]$ at each frame are summed across frequencies as:
\begin{equation}
\mathcal{J}_{\text{SFF-env}}[\tau(\theta),b] = \sum\limits_{k=1}^{K_{SFF}}\mathcal{J}_{\text{SFF-env}}[\tau(\theta),k,b].
\end{equation}
The location of the peak in $\mathcal{J}_{\text{SFF-env}}[\tau(\theta),b]$ at each time frame is estimated as:
\begin{equation}
    \hat{\theta}(b) =  {\argmax_{\theta}}~~ \mathcal{J}_{\text{SFF-env}}[\tau(\theta),b].
\end{equation}
The frame-specific DoA estimates $\hat{\theta}(b)$ are used to obtain a histogram. The peak locations in the histogram correspond to the source DoA estimates as described in Section~\ref{subsec:SFF-PHAT-env}.


\subsection{\label{subsec:descriptionSffphase}SFF-PHAT}
While the DoA estimators like SFF-mean, SFF-var and SFF-env exploit amplitude envelopes of the SFF outputs, SFF-PHAT exploits the phase of SFF outputs and is proposed in \cite{interspeech2020_sffphase}. The idea behind SFF-PHAT DoA estimator is to get better DoA estimates by using only the phase information as in the case of standard GCC-PHAT. While GCC-PHAT is performed in STFT domain, SFF-PHAT is performed in SFF domain.
\par\indent In Figure~\ref{fig:all_sff}(d), a block diagram implementation of SFF-env is shown. If $y_{1}[k,n]$ and $y_{2}[k,n]$ correspond to the $k^{th}$ component of the complex SFF output from two microphone signals at time instant $n$, then the cross correlation function obtained at each time instant is:  
\begin{equation}
   \label{eq:SFFphase1}
    \mathcal{J}_{\text{SFF-PHAT}}[\tau(\theta),n] = \sum_{k=1}^{K_{SFF}}{\frac{y_{1}[k,n]y^*_{2}[k,n]}{|y_{1}[k,n]||y^*_{2}[k,n]|}e^{-j2\pi f_k \tau(\theta)}},
\end{equation}
where $f_k$ is the frequency (in Hertz) corresponding to the $k^{th}$ bin.
The location of the peak in $\mathcal{J}_{\text{SFF-PHAT}}[\tau(\theta),n]$ at each time instant is estimated as:
\begin{equation}
   \label{eq:SFFphase2}
    \hat{\theta}(n) =  {\argmax_{\theta}}~~\mathcal{J}_{\text{SFF-PHAT}}[\tau(\theta),n].
\end{equation}
The instantaneous DoA estimates $\hat{\theta}(n)$ are used to obtain a histogram from which source DoA estimates are obtained as described in Section~\ref{subsec:SFF-PHAT-env}.

\par
\begin{figure}[h!]
  \centering
  \includegraphics[scale=0.43]{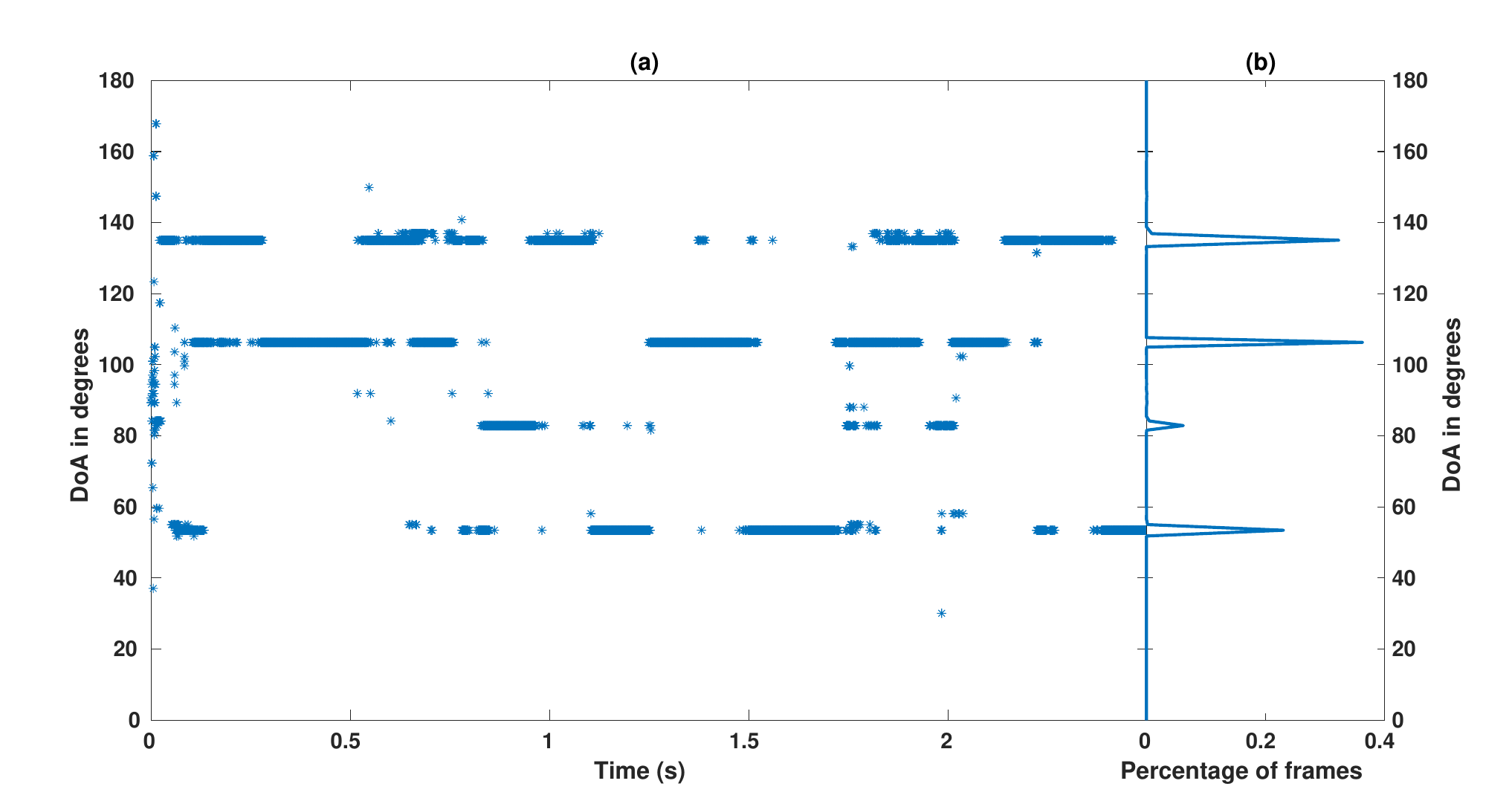}
  \caption[SFF-PHAT DoA estimator applied on a mixture from SiSEC database]{SFF-PHAT DoA estimator applied on a mixture from SiSEC database with 4 female speakers and reverberation time RT60 = 250~ms. $(r)$ Angular spectrogram - shows the DoA estimates of active speakers per each instant. The DoA estimates obtained from histogram of instantaneous DoA estimates in $(r)$ is shown in $(b)$, which is called averaged angular spectrogram.}
  \label{fig:SiSEC_fem4_250ms_sff_phase}
\end{figure}

\indent An illustration of the method of using frame wise $\hat{\theta}(b)$ or instantaneous DoA estimates $\hat{\theta}(n)$ to obtain the source DoA estimates is in Figure~\ref{fig:SiSEC_fem4_250ms_sff_phase}. A stereo recording from SiSEC database (described in Section~\ref{subsec:data} \cite{sisecDetails}) is used for this purpose. The specific recording chosen is that of 4 female concurrent speakers in a room with reverberation time (RT60 = 250~ms) collected at microphones 1m apart. For Illustration, SFF-PHAT DoA estimator is chosen. However, any other SFF-based DoA estimator can be used. Figure~\ref{fig:SiSEC_fem4_250ms_sff_phase} (a) shows the DoA estimates of active speakers at each instant. The histogram of instantaneous DoA estimates is shown in Figure~\ref{fig:SiSEC_fem4_250ms_sff_phase} (b). The percentage of frames in which a speaker is detected is evident from the strength of the histogram peaks. It is observed from  Figure~\ref{fig:SiSEC_fem4_250ms_sff_phase} (b) that the speaker at $80 ^ {o} $ is detected in about $6$ percent of total instants, while the speaker at $105 ^ {o} $ is detected in about $36$ percent of total instants.

\par
\section{\label{sec:Experiments and Results}Experimental Evaluations}
This section gives the details of data sets, baseline DoA estimators, parameters chosen for the DoA estimators, evaluation metrics  and experiments conducted. 

\subsection{\label{subsec:data}Data} This study uses two publicly available data sets consisting of real recordings - Signal Separation and Evaluation Campaign (SiSEC) \cite{sisecDetails} and Challenge on acoustic source LOCalization And TrAcking (LOCATA) \cite{locata}.

\subsubsection{\label{subsubsec:Sisec}SiSEC} The development data of SiSEC $(\textit{dev1}$ and $\textit{dev2}$), which consists ten under-determined stereo channel speech mixtures, is used. To generate the mixtures, first, each static source played through loudspeakers in a meeting room of dimensions 4.45m $\times$ 3.55m $\times$ 2.5m is recorded by omnidirectional, 1m apart pair of microphones. Subsequently, the static source recordings are added to generate the mixtures. The recordings are available with two reverberation times of 130~ms and 250~ms. The DoA of speakers varies between $30^{\circ}$ and $150^{\circ}$ with respect to the array axis. The speakers are 1.2m away from the array. Each mixture is 10s long and consists of 3 or 4 concurrently speaking speakers. These mixtures contain measurement and background noise because they are generated by mixing live recordings from a real room environment. The ground truth values of  DoAs of sources are available for $\textit{dev1}$ and $\textit{dev2}$. In this study, the mixture data considered is 2s instead of 10s. \\
\\ 
\subsubsection{\label{subsubsec:locata}LOCATA} The two tasks in LOCATA that are considered in this study are \textit{Task1}: DoA estimation of a single, static speaker using a static microphone array and \textit{Task2}: DoA estimation of multiple static speakers using a static microphone array. Simultaneous recordings of static sources are collected at multiple microphone arrays in a real-world room environment. Static sources are the loudspeakers from which sentences selected from the CSTR VCTK database~\cite{cstr_vctk} are played back and the room is a computer laboratory with dimensions: 7.1m $\times$ 9.8m $\times$ 3m and $RT60=550\text{~ms}$. These signals are recorded by a non-uniformly spaced linear array of 13 microphones. In this study, a sub-array of two microphones with a separation of 1m is chosen. \textit{Task1} has three recordings, and \textit{Task2} has three recordings of two to four speakers, and the recordings are of duration 3s to 7s. The ground truth values of DoAs of sources are provided. This data set is useful for bench-marking various DoA estimators since the data is generated from live recordings in a highly reverberant room.

\subsection{\label{subsec:Compare_approaches}Baseline DoA Estimators} In this section, the baseline DoA estimators that are compared with the SFF-based DoA estimators are summarized. SFF-based DoA estimators exploit, primarily, excitation source features for DoA estimation. Hence, one of the comparison estimators chosen exploits excitation source features for DoA estimation, and it cross correlates the microphone signals after extracting the epoch locations. This estimator called Hilbert Envelope of Linear Prediction residual (HE-LP) is proposed in \cite{2005_Hilbert_yegna} and is explained in Section~\ref{subsub:HE}. The other comparison estimators are variants of state-of-the-art GCC-based DoA estimators. Two of these are broadband estimators (GCC and GCC-PHAT), and the third one is a narrowband version of GCC called NarrowBand Steered Response Power with PHAT weighting (NB-SRP-PHAT). These estimators are described from Section~\ref{subsub:GCC} to Section~\ref{subsub:SRPPhat}.
\subsubsection{\label{subsub:HE}Hilbert Envelope of Linear Prediction Residual} In HE-LP \cite{2005_Hilbert_yegna}, the excitation source information is obtained from the microphone signals using LP analysis \cite{LP_1975}. Hilbert envelopes of the LP residual of the microphone signals are cross correlated to find the DoA estimates. Hilbert envelope of LP residual signal $\textit{l}[n]$ is given by:
\begin{equation}
    h[n] = \sqrt{\textit{l}^2[n] + \textit{l}_h^2[n]},
\end{equation}
where $\textit{l}_{h}[n]$ is the Hilbert transform of $\textit{l}[n]$.
Let $h_1[n]$ and $h_2[n]$ be Hilbert envelopes of LP residuals of the signals at two microphones. Cross correlation is usually estimated on short frames of Hilbert envelopes of length between 50~ms to 500~ms \cite{2005_Hilbert_yegna}. Let $h_{1b}[n]$ and $h_{2b}[n]$ be two corresponding frames each of length $N$ samples. Then the cross correlation function of these frames is similar to Equation~\eqref{eqn:cssp} and is given by: 

\begin{equation}
	\label{eqn:hilbert}
	\mathcal{J}_{\text{HE-LP}}[\tau(\theta),b] ={\sum\limits_{n=0}^{{N}-1}}{h_{1b}[n]h_{2b}[n+\tau(\theta)]}, \qquad 
	-T_{max}\leq \tau(\theta) \leq T_{max},
\end{equation}
where $T_{max}$ is the maximum time-delay between the given pair of microphones. The cross correlation function is estimated at each frame $b$ where $0\leq b \leq B$. Each block begins at $b\Delta T$. The location of the peak in $\mathcal{J}_{\text{HE-LP}}[\theta,b]$ at each time frame is estimated as:
\begin{equation}
    \hat{\theta}(b) =  {\argmax_{\theta}}~~ \mathcal{J}_{\text{HE-LP}}[\tau(\theta),b].
\end{equation}
The frame specific DoA estimates $\hat{\theta}(b)$ are used to obtain a histogram. The peak locations in the histogram correspond to the source DoA estimates as described in Section~\ref{subsubsec:SFF-mean}.

\subsubsection{\label{subsub:GCC}Generalized Cross Correlation (GCC)}
The cross correlation of filtered versions of the signals  is called Generalized Cross Correlation function \cite{1976_GCC_knapp}.  \C{The cross correlation function $\mathcal{J}_{cc}[\tau(\theta),b]$ obtained by frequency domain operations is in Equation~\eqref{eq:gcc_freq}.} The GCC function $\mathcal{J}_{GCC}[\tau(\theta),b]$ is given by:

\begin{equation}
   \label{eq:gcc_freq}
    \mathcal{J}_{GCC}[\tau(\theta),b] =   \sum_{k}{W(k){X_{1}[k,b]X^*_{2}[k,b]}e^{-j2\pi f_k \tau(\theta)}},
\end{equation}
where $W(k)$ is the weight function, $X_{1}[k,b]$ and $X_{2}[k,b]$ are the STFT of $x_1[n]$ and $x_2[n]$ respectively and  $f_k$ is the frequency (in Hertz) corresponding to the $k^{th}$ bin.  Different weight functions are investigated in~\cite{1976_GCC_knapp}. \C{If $W(k) = 1$, then GCC in Equation~\eqref{eq:gcc_freq} is equivalent to cross correlation function in Equation~\eqref{eq:cc_freq}.} In this study, we consider two variants of GCC as baseline DoA estimators. One with $W(k)=1$ and referred to as GCC and the other is PHAT weighting, which is described in Section~\ref{subsub:GCCPHAT} and referred to as GCC-PHAT. At each time frame $b$, the peak location in $\mathcal{J}_{\text{GCC}}[\tau(\theta),b]$ is estimated, indicating the DoA of an active speaker in that time frame as: 
 \begin{equation}
   \label{eq:bb_gcc_phat2}
    \hat{\theta}(b) =  {\argmax_{\theta}}~~ \mathcal{J}_{\text{GCC}}[\tau(\theta),b].
\end{equation}
The frame specific DoA estimates $\hat{\theta}(b)$ are used to obtain a histogram. The peak locations in the histogram correspond to the source DoA estimates as described in Section~\ref{subsubsec:SFF-mean}.

\subsubsection{\label{subsub:GCCPHAT}Generalized Cross Correlation with Phase Transform Weighting (GCC-PHAT)}
 The signals obtained  at two microphones, $x_1[n]$ and $x_2[n]$, are correlated using GCC-PHAT to obtain the cross correlation function $\mathcal{J}_{\text{GCC-PHAT}}[\tau(\theta),b]$ \cite{1976_GCC_knapp}:

\begin{equation}
   \label{eq:bb_gcc_phat}
    \mathcal{J}_{\text{GCC-PHAT}}[\tau(\theta),b] =   \text{GCC-PHAT}(x_1[n],x_2[n]) =\sum_{k}{\frac{X_{1}[k,b]X^*_{2}[k,b]}{|X_{1}[k,b]||X^*_{2}[k,b]|}e^{-j2\pi f_k \tau(\theta)}},
\end{equation}
where $X_{1}[k,b]$ and $X_{2}[k,b]$ are the STFT of $x_1[n]$ and $x_2[n]$ respectively and $f_k$ is the frequency (in Hertz) corresponding to the $k^{th}$ bin.
At each time frame $b$, the peak location in $\mathcal{J}_{\text{GCC-PHAT}}[\tau(\theta),b]$ is estimated, indicating the DoA of an active speaker in that time frame as: 
 \begin{equation}
   \label{eq:bb_gcc_phat2}
    \hat{\theta}(b) =  {\argmax_{\theta}}~~ \mathcal{J}_{\text{GCC-PHAT}}[\tau(\theta),b].
\end{equation}
The frame specific DoA estimates $\hat{\theta}(b)$ are used to obtain a histogram. The peak locations in the histogram correspond to the source DoA estimates as described in Section~\ref{subsubsec:SFF-mean}. GCC-PHAT may also be referred to as BroadBand Steered Response Power with PHAT weighting (BB-SRP-PHAT).
\subsubsection{\label{subsub:SRPPhat}Narrowband Steered Response Power with Phase Transform Weighting (NB-SRP-PHAT)}
In broadband estimators such as GCC and GCC-PHAT, a DoA estimate is obtained in each frame. Whereas in narrowband estimators, a DoA estimate is obtained for each TF bin \cite{madhu08mog}. In both broadband and narrowband estimators, the local DoAs are subsequently clustered to obtain the target DoA. In NB-SRP-PHAT, the target DoA is found from histogram of the local DoA estimates at each TF bin. \par\indent The DoA estimate is obtained at a time frame $b$ and frequency bin $k$ as follows:
\begin{equation}
   \label{eq:NB-SRP-PHAT}
    \mathcal{J}_{\text{NB-SRP-PHAT}}[\tau(\theta),k,b] =   \frac{X_{1}[k,b]X^*_{2}[k,b]}{|X_{1}[k,b]||X^*_{2}[k,b]|}e^{-j2\pi f_k\tau(\theta)}.
\end{equation}
At each TF bin, the peak location in $\mathcal{J}_{\text{NB-SRP-PHAT}}[\tau(\theta),k,b]$ is estimated, indicating the DoA of an active speaker in that TF bin as: 
 \begin{equation}
   \label{eq:NB-SRP-PHAT2}
    \hat{\theta}(k,b) =  {\argmax_{\theta}}~~ \mathcal{J}_{\text{NB-SRP-PHAT}}[\tau(\theta),k,b].
\end{equation}
A histogram of DoA estimates at each TF bin ($\hat{\theta}(k,b)$) is obtained. The locations of the peaks in the histogram are the required source DoA estimates.

\subsection{\label{subsec:evalmetrics}Evaluation Metrics}
The evaluation metrics chosen are two detection metrics: (i) percentage of frames in which speakers are detected ($\alpha$) and (ii) F-measure, and two location accuracy metrics: (i) mean azimuth error (MAE) and (ii) MAEfine. $\alpha$ is obtained from histogram plots by dividing the number of frames in which a source is detected to the total number of frames as mentioned in Equation~\ref{eq:alpha} and is obtained as:
\begin{equation}
    \alpha = \frac{no.\ of\ frames\ in\ which\ a\ speaker\ is\ correctly\ estimated}{total\ no.\ of\ frames}.
    \label{eq:alpha}
\end{equation} Higher the value of $\alpha$, better are the tracking abilities of the DoA estimator. F-measure is defined as follows: In a given data set, let $N_T$ be the total number of sources in all mixture files and $N_e$ be the number of sources that are estimated by an estimator. By greedy matching algorithm, the estimated DoAs for each mixture are  matched to the ground truth DoAs to ensure minimum azimuth error. The source is said to be {\em correctly} estimated if after matching, the estimated source DoA is within $\pm 5^{\circ}$ of the ground truth DoA. Let $N_c$ be the number of sources correctly estimated for all mixtures. Then the F-measure is given by:
\begin{equation}
\text{F-measure} = \dfrac{2*\text{Recall}*\text{Precision}}{\text{Recall}+\text{Precision}},
\end{equation}
where, $\text{Recall} = {N_c}/{N_T}$ and $\text{Precision} ={N_c}/{N_e} $.
The more the number of sources correctly estimated, the higher the F-measure. To quantify the DoA estimation accuracy, two error metrics are reported: MAE and MAEfine. MAE is the mean azimuth error between the estimated DoAs and true DoAs after greedy matching, and MAEfine is the mean error between the {\em correctly} estimated DoAs and true DoAs. Thus, while MAE gives estimation accuracy over all the sources in the mixture, MAEfine gives estimation accuracy of only the {\em correctly} detected sources. Therefore, MAE may be considered as a global performance metric, whereas MAEfine is a local performance criterion with respect to correctly detected sources. For a better performing DoA estimator, F-measure and $\alpha$ would be high, and MAE and MAEfine would be low.

\subsection{\label{subsec:eval_setup}Parameter Setting} In order to fairly compare the SFF-based DoA estimators with the GCC-based DoA estimators, it is necessary to set the parameters to obtain TF representations of SFF and STFT methods. That is, for example, it is not correct to use a small number of frequency bins for STFT analysis and a large number for SFF analysis. It has to be ensured that the window sizes and other parameters are chosen in a systematic way. To attain these comparable parameters, it is useful to interpret STFT as a $\textit{filtering}$ operation. Implementation of STFT as a \textit{filtering} operation is presented in Section~\ref{subsubsec:stftasfilter}. In Section~\ref{subsubsec:link_stft_sff}, the method to obtain comparable parameters for SFF and STFT TF representations is presented. 
\subsubsection{\label{subsubsec:stftasfilter} Short-Time Fourier Transform of Speech Signals by Filtering}
The STFT of a causal sequence $x[n]$ is a sequence of Fourier transforms of the windowed $x[n]$ and is given by:
 \begin{equation}
 \label{eq_dft_eq1}
     X(k,b) = \sum_{n'=0}^{\infty}(x[n']f_{stft}(n'-b\Delta T))e^{-j\omega_k n'},
 \end{equation}
where $f_{\text{STFT}}[n]$ is the STFT analysis window and is usually chosen to have finite temporal duration of $N$ samples and low-pass frequency response.  $0 \leq k \leq K-1$ and $0 \leq b \leq B-1$, where $K$ is the number of DTF points and $B$ is the number of time frames. Due to conjugate symmetry of the discrete Fourier transforms for real-valued signals, only $\frac{K}{2}+1$ frequency bins are used. The $b^{th}$ time frame begins at the sample index given by $b\Delta T$, where $\Delta T$ is the STFT hop size in samples.  In this study, Hann window is used. \par\indent Apart from the common procedure of obtaining STFT representation of speech signals that is described above, another equivalent procedure to get STFT representation is by using a \textit{filter} interpretation as follows: the frequency shifted signal $x[n']e^{-j\omega_k n'}$ is passed through a time reversed (or flipped) filter with impulse response $f_{\text{STFT}}(n)$. The frequency is shifted down to 0~Hz by modulating with $e^{-j\omega_k n}$ in the time domain. This interpretation of STFT in mathematical form is (equal to Equation~\eqref{eq_dft_eq1}):
 \begin{equation}
 \label{eq_dft_eq2}
     X(k,b) =  \sum_{n'=0}^{\infty}(x[n']e^{-j\omega_k n'})f_{\text{STFT}}(n'-b\Delta T) = (x[m]e^{-j\omega_k m}) \mathbf{*} Flip(f_{\text{STFT}}(m)),
 \end{equation}
where $m=b\Delta T$, $\mathbf{*}$ is convolution operation and $f_{\text{STFT}}(n)$ is now viewed as the impulse response of an FIR filter with filter length $N$. $Flip$ is the time-reversed version of the filter and for a symmetrical window $Flip(f_{\text{STFT}}(n)) = f_{\text{STFT}}(n)$. As the value of $N$ increases, the 3~dB bandwidth of frequency response ${B_{STFT}}$ decreases. For a Hann window, the filter length $N$ (in samples) and the $3~dB$ bandwidth of the filter ${B_{STFT}}$ (in $Hz$) are inversely related by ${B_{STFT}} = \frac{1.44f_s}{N}$ \cite{Heinzel2002SpectrumAS}. \par\indent While in SFF, $x[n]$ is frequency shifted to $\frac{f_s}{2}$ Hz (by multiplying with a complex exponential) and passed through a filter with infinite impulse response  $f_{\text{SFF}}[n] = (-r)^n u[n]$, in STFT, the signal frequency shifted to $0$ Hz and passed through a filter with finite impulse response of $N$ samples (Hann function). In SFF, outputs of the filter are obtained at $K_{SFF}$ frequencies lying between $0$ and $\frac{f_s}{2}$. Similarly, the STFT outputs may be obtained at $K_{STFT}$ frequencies lying between $0$ and $\frac{f_s}{2}$.
\subsubsection{\label{subsubsec:link_stft_sff}Method Adapted to get SFF and STFT parameters}
In this section, the procedures adapted for setting the parameters of nine DoA estimators are described. Of the nine DoA estimators, 5 are implemented in SFF TF representation (SFF-PHAT-env, SFF-mean, SFF-var, SFF-PHAT, SFF-env), 3 are implemented in STFT TF representation (GCC, GCC-PHAT, NB-SRP-PHAT) and 1 is implemented in time domain (HE-LP). For a fair comparison, SFF and STFT TF representations are obtained such that the band width of the respective filters and the number of frequency bins in each representation are the same. Table~\ref{tab:summary_param} summarizes the relations between the window parameters and the 3~dB bandwidths in SFF and STFT TF representations. 
\begin{table}[h!]
\small
\centering
\caption[Method to get equivalent SFF and STFT parameters]{Method to get equivalent SFF and STFT parameters. The parameters $r$ and $N$ are set such that ${B_{SFF}}$=${B_{STFT}}$, and $K_{SFF}$ is set equal to $K_{STFT}$.}
\label{tab:summary_param}
\begin{tabular}{|l|l|l|}
\hline
& {SFF}    & STFT                                                                         \\ \hline

{Parameters in the TF representation}                                                                                          & $r$ and $K_{SFF}$                                                                                                    & $N$ and $K_{STFT}$                                                                                                               \\ \hline
{\begin{tabular}[c]{@{}l@{}}Relation between time domain filter parameters \\ ($r \& N$)  and the respective frequency  \\ domain parameters (${B_{SFF}} \& {B_{STFT}}$)
\end{tabular}} & 
${B_{SFF}} = \frac{f_s}{2\pi} \cos^{-1}\big(\frac{4r-r^2-1}{2r}\big)$~\cite{2019_icassp_krishna}                                                                                                                  &   ${B_{STFT}} = \frac{1.44f_s}{N}$~\cite{Heinzel2002SpectrumAS}                                                                                                                           \\ \hline
\end{tabular}
\end{table}
\par\noindent\textbf{Setting of $K_{SFF}$ and $K_{STFT}$:} To choose suitable values of $K_{SFF}$ and $K_{STFT}$, we test the performance of a SFF-based and GCC-based methods for various values of $K_{SFF}$ ($128 \leq K_{SFF} \leq 512$) with $r$ and $N$ fixed and choose the value resulting in best metrics. We fixed the upper limit of $K_{SFF}$ and $K_{STFT}$ to $512$ to keep the computational complexity low. As $K_{SFF}$ and $K_{STFT}$ increase, computational complexity in terms of computation time and memory requirements also increases. Specifically, $K_{SFF}$ and $K_{STFT}$ are chosen based on the performance (F-measure, $\alpha$, MAE) of various DoA estimators on LOCATA data. It is observed that the value of MAEfine does not change significantly with varying $K_{SFF/STFT}$. Hence, MAEfine metric is not used for performance evaluation. Table~\ref{tab:Kparams} gives the performance of SFF-based and GCC-based DoA estimators for different values of $K_{SFF}$/$K_{STFT}$ with $r = 0.99887$ and $N=8000~(500~ms)$ respectively corresponding to 3~dB bandwidth of $\approx 3~Hz.$
\begin{table}[h!]
\caption[Table to choose a suitable value for $K_{SFF}$ and $K_{STFT}$.] {Table to choose a suitable value for $K_{SFF}$ and $K_{STFT}$. F-measure, MAE, and $\alpha$ for various DoA estimators for varying $K_{SFF}$ and $K_{STFT}$ on LOCATA. The 3~dB bandwidth of the filters used is 3~Hz. The best metrics for each DoA estimator among different $K$ are highlighted in \textbf{bold}.}
\label{tab:Kparams}
\begin{tabular}{|l||l|l|l||l|l|l||l|l|l|}
\hline
\textbf{$K_{SFF}$=$K_{STFT}$} & \multicolumn{3}{c||}{{128}} & \multicolumn{3}{c||}{{256}} & \multicolumn{3}{c|}{{512}} \\ \hline
\multicolumn{10}{|c|}{{SFF-based DoA estimators (r = 0.99887 \& $B_{SFF}=3~Hz$)}}                                                                                 \\ \hline
 & F-mea&MAE&$\alpha$&F-mea&MAE&$\alpha$&F-mea&MAE&$\alpha$ \\ 
 & -sure&&&-sure&&&-sure&&\\\hline
{SFF-env}&\textbf{ 0.48}&\textbf{10.07}& 27
& 0.38& 10.75& 30                                                                                                 & 0.38& 14.81& \textbf{28}  \\ \hline
{SFF-PHAT-env}  & \textbf{0.40}& 10.99& 33                                                                                                                 & \textbf{0.40}& 10.99& \textbf{35}                                                                                                                 & \textbf{0.40}& \textbf{10.77}&\textbf{35}  \\ \hline
{SFF-PHAT} & \textbf{0.40}& \textbf{10.65}& 36                                                                                                                & \textbf{0.40}& \textbf{10.65}& 39                                                                                                                & \textbf{0.40}& \textbf{10.65}&\textbf{42}                                                                                                               \\ \hline
{SFF-mean}     & \textbf{0.38}& 15.39& 24                                                                                                                & \textbf{0.38}& 14.25& 29                                                                                                                & \textbf{0.38}& \textbf{10.88}&\textbf{33}                                                                                                                \\ \hline
{SFF-var}      & 0.31& 22.82& 19                                                                                                                & 0.38& 14.60& 22                                                                                                                & \textbf{0.46}& \textbf{11.69}& \textbf{28}                                                                                                                \\ \hline
\multicolumn{10}{|c|}{{GCC-based DoA estimators~(N=8000~(500~ms) \& $B_{STFT}=3~Hz$)}}                                                                                  \\ \hline
                              
{GCC}          & 0.38& 14.75& 29                                                                                                                & 0.38& 13.80& 35                                                                                                                & \textbf{0.46}& \textbf{12.69}&\textbf{ 36}                                                                                                                \\ \hline
{GCC-PHAT}     & 0.38& \textbf{10.65}& 44                                                                                                                & 0.38& \textbf{10.65}& 45                                                                                                                & \textbf{0.46}& 12.44& \textbf{46}                                                                                                                \\ \hline
{NB-SRP-PHAT}  & \textbf{0.38}& 19.7& NA                                                                                                                 & 0.31& 17.34& NA                                                                                                                & \textbf{0.38}& \textbf{14.60}& NA  \\ \hline
\end{tabular}
\end{table}


From the table, it is observed that in most of the cases the performance metrics increase or remain constant with increasing $K$ values. The best metrics for each DoA estimator are highlighted in \textbf{bold} and are obtained at $512$. Hence we chose  
$K_{SFF}=K_{STFT}=512$. 

\textbf{Setting of ${B_{SFF}}$ and ${B_{STFT}}$:} In literature, for TDoA estimation of signals at two microphones, the signals are divided into frames of size of a few hundred milliseconds (ranging from 50~ms to 500~ms) \cite{2005_Hilbert_yegna, dibiase}. The choice of the frame size is dictated by the application-while small frame size is suitable for tracking, a large frame size would improve the performance of the TDoA estimators. In this study, to choose an appropriate frame size, we test all the baseline estimators with varying frame sizes (N varying from 64~ms to 500~ms) on LOCATA data. For every N, we find a corresponding $r$, such that ${B_{SFF}} = {B_{STFT}}$ and test the performance of various SFF-based estimators. In Table~\ref{tab:bwparams}, the performance metrics of various DoA estimators with increasing $N$ and the corresponding $r$ are shown. It is observed from the table that the performance of the estimators improve as the $r$ and $N$ increases and it is inline with the theory that increasing frame size lead to improved DoA estimates. The best metrics are obtained for the case of $N=8000$ and $r=0.99887$. Hence in the following experiments we fix $N=8000$ and $r=0.99887$.  
\begin{table}[h!]
\small
\caption[Table to choose suitable values for $r$ and $N$.] {Table to choose suitable values for $r$ and $N$. F-measure, MAE, and $\alpha$ for various DoA estimators for varying $r$ and $N$ on LOCATA. The values of $r$ and $N$ are chosen such that, $r$ and $N$ in the same column result in equal 3~dB bandwidth of SFF and STFT filters (${B_{SFF}}$ = ${B_{STFT}}$). $K_{SFF}=K_{STFT}=512$. The best metrics among the different $r$ and $N$ for each DoA estimator are highlighted in \textbf{bold}.}
\label{tab:bwparams}
\begin{tabular}{|l||l|l|l||l|l|l||l|l|l||l|l|l|}

\hline
\multicolumn{13}{|c|}{{SFF-based DoA estimators}($K_{SFF}=512$)}  \\ \hline
~~~~~~~~{r~=~} & \multicolumn{3}{c||}{{0.99125}} & \multicolumn{3}{c||}{{0.99718}} & \multicolumn{3}{c|}{{0.99859}} &
\multicolumn{3}{c|}{{0.99887}}\\ \hline                                                                            
 & F-mea&MAE&$\alpha$&F-mea&MAE&$\alpha$&F-mea&MAE&$\alpha$&F-mea&MAE&$\alpha$ \\ 
 & -sure&&&-sure&&&-sure&&&-sure&&\\\hline
\scriptsize{{SFF-env}}      & \textbf{0.38}& \textbf{10.35}& 17                                                                                                            & \textbf{0.38}& 10.86& 21                                                                                                             & \textbf{0.38}& 15.33& 27                                                                                                             & \textbf{0.38}& 14.81& \textbf{28}                                                                                                             \\ \hline
\scriptsize{{SFF-PHAT-env}} & 0.38& 11.05& 20                                                                                                            & 0.38& 10.9& 28                                                                                                              & \textbf{0.40}& 10.99& 33                                                                                                              & \textbf{0.40}&\textbf{ 10.77}&\textbf{ 35}                                                                                                              \\ \hline
\scriptsize{{SFF-PHAT}}    & \textbf{0.40}& \textbf{10.65}& 14                                                                                                            & \textbf{0.40}&\textbf{ 10.65}& 30                                                                                                             & \textbf{0.40}&\textbf{ 10.65}& 39                                                                                                             & \textbf{0.40}&\textbf{ 10.65}& \textbf{42}                                                                                                             \\ \hline
\scriptsize{{SFF-mean}}     & \textbf{0.38}& \textbf{10.69}& 15                                                                                                            & \textbf{0.38}&\textbf{ 10.69}& 22                                                                                                             & \textbf{0.38}& 11.26& 29                                                                                                             & \textbf{0.38}& \textbf{10.88}& \textbf{33}                                                                                                             \\ \hline
\scriptsize{{SFF-var}}      & 0.46& 13.79& 14                                                                                                            & 0.38& 11.26& 22                                                                                                             & 0.38& 15.19& 26                                                                                                             & \textbf{0.46}& \textbf{11.69}& \textbf{28}                                                                                                             \\ \hline
 \multicolumn{13}{|c|}{{GCC-based DoA estimators}($K_{STFT}=512$)}\\ \hline             
 ~~~~~~~~{N~=~} & \multicolumn{3}{c||}{{1024(64~ms)}} & \multicolumn{3}{c||}{{3200(200~ms)}} & \multicolumn{3}{c|}{{6400(400~ms)}} &
\multicolumn{3}{c|}{{8000(500~ms)}}\\ \hline  
\scriptsize{{GCC}}          & 0.38& \textbf{10.77}& 21                                                                                                            & 0.38&\textbf{10.77}& 28                                                                                                              & \textbf{0.46}& 12.68& 34                                                                                                             & \textbf{0.46}& 12.69& \textbf{36}                                                                                                             \\ \hline
\scriptsize{{GCC-PHAT}}     & 0.38& \textbf{10.65}& 35                                                                                                            & 0.38& \textbf{10.65}& 42                                                                                                             & \textbf{0.46}& 12.44& 45                                                                                                             & \textbf{0.46}& 12.44&\textbf{ 46}                                                                                                             \\ \hline
\scriptsize{{NB-SRP-PHAT}}&\textbf{0.38}& 20.8& NA &0.31& 20.7& NA&0.31& 25& NA&\textbf{0.38}&  \textbf{14.60}& NA \\ \hline                                                   
\end{tabular}
\end{table}

\par\indent Further, in HE-LP (LP residual is obtained on $20~ms$ frame) and SFF-env, cross-correlation is performed with frame size of 8000 samples. In the SFF-based DoA estimators, after obtaining the SFF TF representation of the microphone signals, GCC-PHAT of either the envelopes (as in SFF-\textit{PHAT-env}) or the spectral features (as in SFF-mean, SFF-var) is performed. In these cases, STFT is obtained with the frame size = 8000 samples, frame shift = 80 samples and DFT points = 8000. From the 4000 bins corresponding to frequencies $0$ to $\frac{f_s}{2}$, equally spaced 512 bins are used for GCC-PHAT.          


\subsection{\label{subsec:Experiment set-up}Experimental Details} 
In all the experiments, stereo microphones with inter microphone spacing of $1~m$ are considered. The recordings used for the experiments are captured at the microphones in real room acoustic environments as described in Section~\ref{subsec:data}. Therefore, the microphone signals in SiSEC and LOCATA are corrupted by room reverberation and background noise. Further, the SiSEC microphone data is corrupted by adding different types of uncorrelated noises (white, pink, machine gun, volvo, and babble) at SNR of $0~dB$ taken from NOISEX database~\cite{Noisex}.

\section{\label{sec:ResultsAndDiscussion}Results and Discussion}
This section presents the results of the experiments described in Section~\ref{subsec:Experiment set-up}, along with a discussion. In Section~\ref{subsub:sisec&locata}, performance comparison of SFF-based (SFF-env, SFF-\textit{PHAT-env}, SFF-PHAT, SFF-mean and SFF-var) and baseline DoA (GCC, GCC-PHAT, NB-SRP-PHAT and HE-LP) estimators on SiSEC and LOCATA data sets is presented in terms of the evaluation metrics discussed in Section~\ref{subsec:evalmetrics}. In addition, to study the effect of DoA estimators on noisy speech, the DoA estimators are tested on SiSEC recordings corrupted with different types of noises in Section~\ref{subsub:noisySisec}. Also, a graphical illustration of the performance of various DoA estimators on SiSEC and noisy SiSEC data is presented for one randomly chosen recording. A  \textit{weighted} GCC-PHAT DoA estimator is implemented and tested on the noisy data with the idea of comparing its performance with the best performing SFF-based estimators.\\

\subsection{\label{subsub:sisec&locata}Results and Discussion of Experiments Conducted on SiSEC and LOCATA Data}
Table~\ref{tab:SiSEC&Locata} gives the results of SFF-based (SFF-env, SFF-\textit{PHAT-env}, SFF-PHAT, SFF-mean and SFF-var) and baseline DoA estimators (GCC, GCC-PHAT, NB-SRP-PHAT and HE-LP) on SiSEC and LOCATA data sets. Except in NB-SRP-PHAT, in all the estimators, the position of r single peak in the correlation function (indicating dominant speaker per frame) is detected in each frame for calculating $\alpha$. But for NB-SRP-PHAT, a speaker may be detected multiple times in a single frame. Therefore, $\alpha$ is not reported for NB-SRP-PHAT. \\ 
\begin{table}[h!]

\caption[Performance comparison of SFF-based and baseline DoA estimators on SiSEC and LOCATA.]{\label{tab:SiSEC&Locata} {Performance comparison of SFF-based (SFF-env, SFF-\textit{PHAT-env}, SFF-PHAT, SFF-mean and SFF-var) and baseline (GCC, GCC-PHAT, NB-SRP-PHAT and HE-LP) DoA estimators in terms of F-measure, MAE, MAEfine and $\alpha$ on SiSEC and LOCATA data. The two best metrics among various estimators are highlighted in \textbf{bold}.}}
\centering
\begin{tabular}{|l||l|l|l|l|l||l|l|l|l|}
\hline
& \multicolumn{5}{c||}{{SFF - based DoA estimators}}&\multicolumn{4}{c|}{{Baseline DoA estimators}}\\
\hline
 & {SFF-} & {SFF-} &  {SFF-} & {SFF-} & {SFF-} &  {GCC} &  {GCC-} &  {NB-} &  {HE-LP} \\
 
 &{env} & {PHAT-} &{PHAT} & {mean} & {var} & & {PHAT}& {SRP-} & \\
 
  & & {env} & & &  & & & {PHAT} & \\
 \hline
 \multicolumn{10}{|c|}{{SiSEC}} \\ \hline 
 F-measure&\textbf{1.00}&\textbf{1.00}&\textbf{1.00}&0.94&{0.81}&0.86&{0.97}&0.86&0.70\\
 \hline
 MAE&1.07&\textbf{0.79}&\textbf{0.78}&3.24&5.24&2.67&1.33&8.28&9.15\\
 \hline
 MAEfine&1.07&\textbf{0.77}&\textbf{0.77}&0.88&1.15&\textbf{0.60}&{0.83}&1.20&\textbf{0.77}\\
 \hline
 $\alpha$&59&87&\textbf{96}&69&58&66&\textbf{97}&NA&50\\
 \hline
 \multicolumn{10}{|c|}{{LOCATA}} \\ \hline
  F-measure &0.38&\textbf{0.40}&\textbf{0.40}&0.38&\textbf{0.46}&\textbf{0.46}&\textbf{0.46}&0.38&\textbf{0.46}\\
 \hline
 MAE & 14.81&\textbf{10.77}&\textbf{10.65}&{10.88}&{11.69}&12.69&12.44&14.60&13.24\\
 \hline
 MAEfine &\textbf{3.33}&3.91&3.56&3.86&\textbf{3.06}&3.52&3.72&4.33&3.65\\
 \hline
 $\alpha$&28&35&\textbf{42}&33&28&36&\textbf{46}&NA&27\\
 \hline
\end{tabular}
\end{table}
\indent From the results on SiSEC (Table~\ref{tab:SiSEC&Locata}), it is clearly observed that among the SFF-based DoA estimators, SFF-PHAT is the best performing estimator in terms of all metrics. It is followed by SFF-PHAT-env. SFF-var is the least performing estimator in terms of all metrics. Among the baseline DoA estimators, GCC-PHAT is the best performing in all the metrics. GCC is the next best. HE-LP and NB-SRP-PHAT perform poorly. Between the best SFF-based and baseline estimators, SFF-PHAT and GCC-PHAT give comparable results and closely followed by SFF-PHAT-env. 

\indent From the results of  LOCATA (Table~\ref{tab:SiSEC&Locata}), the following points are evident. The performance of all the DoA estimators degrades compared to the performance on SiSEC. This is because of the high reverberation time in LOCATA (up to 550~ms) compared to low reverberation time in SiSEC (up to 250~ms). Among the SFF-based DoA estimators, SFF-PHAT is the best performing estimator in terms of all metrics. It is followed by SFF-PHAT-env and SFF-mean. SFF-var and SFF-env are the least performing estimators especially in terms of the detection metric $\alpha$. Among the baseline DoA estimators, GCC-PHAT is the best performing in terms of all metrics. NB-SRP-PHAT performs poorly. Among both SFF-based and baseline estimators, SFF-PHAT and GCC-PHAT are comparable. 
Combining SiSEC and LOCATA results, we make the following observations. The increasing order of performance of SFF-based DoA estimators is as follows: 
\\ {\textsf{ SFF-var $\leq$ SFF-env $<$  SFF-mean $\leq$ SFF-\textit{PHAT-env} $<$ SFF-PHAT}}.\\ 
The increasing order of performance of baseline DoA estimators is as follows:
\\ {\textsf{NB-SRP-PHAT $\leq$ HE-LP $<$ GCC $<$ GCC-PHAT}}.\\ The reason for poor performance of NB-SRP-PHAT is that, it fails to provide a unique maximum for for frequencies above the spatial aliasing frequency. As the inter-microphone distance increases, a larger range of frequencies are affected by spatial aliasing, and the efficacy of NB-SRP-based methods decreases. Between SFF-PHAT and GCC-PHAT, which are the best performing SFF-based and baseline DoA estimators, respectively, the performance of both the estimators is comparable in all metrics. The proposed estimator SFF-PHAT-env is the next best after SFF-PHAT and its performance is significantly better than SFF-env. \C{Finally, the benefit GCC-NMF, which is an NMF weightd GCC-PHAT, is clearly seen from the improved metrics of GCC-NMF than the GCC-PHAT.}



\subsection{\label{subsub:noisySisec}Results of Experiments Conducted on Noisy SiSEC Data}
This section presents the effect of noise on the performance of DoA estimators. Different types of noise (white, babble, pink, volvo and machine gun) are added to SiSEC data at SNR of $0~dB$ such that the noise at the two microphones is uncorrelated. Table~\ref{tab:SiSECNoisy} gives the metrics of SFF-based and baseline DoA estimators, averaged over three sets of simulations, on noisy SiSEC data. Each simulation set consists of ten recordings consisting of 36 speakers.

\begin{table}[h!]
\centering

\caption [Performance comparison of SFF-based and baseline DoA estimators on noisy SiSEC.] {\label{tab:SiSECNoisy} {Performance comparison of SFF-based (SFF-env, SFF-\textit{PHAT-env}, SFF-PHAT, SFF-mean and SFF-var) and baseline (GCC, GCC-PHAT, NB-SRP-PHAT and HE-LP) DoA estimators in terms of F-measure, MAE, MAEfine and $\alpha$ on noisy SiSEC data. The noises (white, babble, pink, volvo and machine gun (M. gun)) are added at SNR = 0~dB. The two best metrics among various DoA estimators are highlighted in \textbf{bold}.}}

\begin{tabular}{|l||l|l|l|l|l||l|l|l|l|}

\hline
& \multicolumn{5}{c||}{{SFF - based DoA estimators}}                                                                                                                                                                    & \multicolumn{4}{c|}{{Baseline DoA estimators}} \\   
 \hline
 & {SFF-} & {SFF-} &  {SFF-} & {SFF-} & {SFF-} &  {GCC} &  {GCC-} &  {NB-} &  {HE-LP} \\
 
 &{env} & {PHAT-} &{PHAT} & {mean} & {var} & & {PHAT}& {SRP-} & \\
 
  & & {env} & & &  & & & {PHAT} & \\
  \hline
 \multicolumn{10}{|c|}{{SiSEC+White}}                                                                                                                                                                                                                                                                                                                                                                                                                                                                                                                        \\ \hline

 F-measure &0.27 & 0.5&\textbf{0.69}&0.31&0.24&0.29&\textbf{0.57}&0.47&0.24\\
 \hline
 MAE & 19.96&{16.28}&\textbf{10.54}&19.89&20.32&19.49&\textbf{11.76}&25.53&21.64\\
 \hline
 MAEfine &2.36&\textbf{1.98}&\textbf{1.62}&2.72&2.42&{2.51}&1.79&2.22&2.45\\
 \hline
 $\alpha$ &4&\textbf{17}&10&9&6&{8}&\textbf{17}&NA&8\\
  \hline
 \multicolumn{10}{|c|}{\textbf{SiSEC+Babble}}                                                                                                                                                                                                                                                                                                                                                                                                                                                                                                                                                                                                                      \\ \hline
 F-measure & 0.78&\textbf{0.91}&\textbf{0.93}&0.54&0.46&0.69&0.74&0.65&0.30\\
 \hline
 MAE & 7.18&\textbf{3.45}&\textbf{1.14}&14.47&17.05&8.33&7.47&16.67&21.61\\
 \hline
 MAEfine &1.56&1.62&\textbf{0.79}&2.01&2.41&\textbf{0.85}&{0.93}&1.54&1.66\\
 \hline
 $\alpha$ &{32}&\textbf{61}&48&24&22&34&\textbf{53}&NA&10\\

   \hline
 \multicolumn{10}{|c|}{\textbf{SiSEC+Pink}}                                                                                                                                                                                                                                                                                                                                                                                                                                                                                                                                                                                                                      \\ \hline
 F-measure &0.47&{0.54}&\textbf{0.74}&0.28&0.31&0.5&\textbf{0.60}&0.46&0.28\\
 \hline
 MAE &16.5&\textbf{11.34}&\textbf{8.03}&19.05&26.95&14.31&\textbf{11.37}&25.64&18.43\\
 \hline
 MAEfine &2.03&1.75&\textbf{1.25}&2.4&2.07&\textbf{1.63}&1.80&2.39&3.11\\
 \hline
 $\alpha$ &13&\textbf{24}&12&8&10&\textbf{18}&17&NA&8\\

  \hline
 \multicolumn{10}{|c|}{\textbf{SiSEC+Volvo}}                                                                                                                                                                                                                                                                                                                                                                                                                                                                                                                                                                                                                      \\ \hline
 F-measure & 0.88&\textbf{0.97} &\textbf{1} &0.90 &0.83 &0.81 &0.90 & 0.85&0.61\\
 \hline
 MAE & 4.24 &\textbf{0.95} &\textbf{0.78} &4.11 &5.43 &2.59 &4.16 &9.12 &14.54\\
 \hline
 MAEfine & 1.18 &0.95 &\textbf{0.78} &1.39 &1.74 &\textbf{0.59} &{0.84} &1.15 &0.88\\
 \hline
 $\alpha$ &48&87&\textbf{93}&60&54&58&\textbf{96}&NA&35\\
  \hline
 \multicolumn{10}{|c|}{\textbf{SiSEC+M.Gun}}                                                                                                                                                                                                                                                                                                                                                                                                                                                                                                                                                                                                                      \\ \hline
 F-measure &0.87 &\textbf{0.97} &\textbf{1.00} &0.86 &0.82 &0.85 &0.86 &0.82 &0.67\\
 \hline
 MAE & 4.19&\textbf{1.03}&\textbf{0.78}&3.71&9.18&2.73&4.04&12.71&8.25\\
 \hline
 MAEfine & 1.31 &1.03 &\textbf{0.78} &1.55 &1.89 &\textbf{0.69} &{0.85} &1.12 &1.09\\
 \hline
 $\alpha$ &44&86&\textbf{92}&51&47&57&\textbf{94}&NA&37\\
 \hline
\end{tabular}
\end{table} 
It is observed from Table~\ref{tab:SiSECNoisy} that the performance of all the DoA estimators has degraded when compared to the performance on clean SiSEC data that is tabulated in Table~\ref{tab:SiSEC&Locata}. SFF-PHAT and SFF-PHAT-env are the best performing SFF-based estimators. GCC-PHAT is the best performing in the baseline estimators. In all the noises, SFF-PHAT has highest F-measure, lowest MAE and lowest MAEfine among all the estimators. Between the proposed SFF-PHAT-env and GCC-PHAT, in babble, volvo and machine gun noises, SFF-PHAT-env is superior in F-measure and MAE metrics and have comparable $\alpha$. In other noises they are comparable in terms of all metrics. Therefore it may stated that SFF-PHAT is the best performing among all the estimators and followed closely by SFF-PHAT-env and GCC-PHAT. The proposed SFF-PHAT-env is superior to SFF-env in all the metrics and in all the cases. HE-LP performs poorly because its performance mainly depends on the performance of linear prediction analysis, which deteriorates in low SNR conditions \cite{1999_enhance_LP_yegna}. The results clearly show that DoA estimation in SFF domain is superior to DoA estimation in the STFT domain. This point is better appreciated by comparing GCC-PHAT and SFF-PHAT. The only difference between these two DoA estimators is the TF representation. While GCC-PHAT is implemented in STFT domain, SFF-PHAT is implemented in SFF domain. Both the methods use phase information to find the DoAs.
Among all the noises, the least performance degradation is observed in volvo and machine gun noises. Among the noises considered, the increasing order of the effect of noise on the performance of the DoA estimators is as follows: \par\indent \indent \indent \indent \indent \textsf{Volvo $<$ Machine gun $<$ Babble $<$ Pink $<$ White}.\par
\begin{figure}[h!]
  \centering
  \includegraphics[scale=0.45]{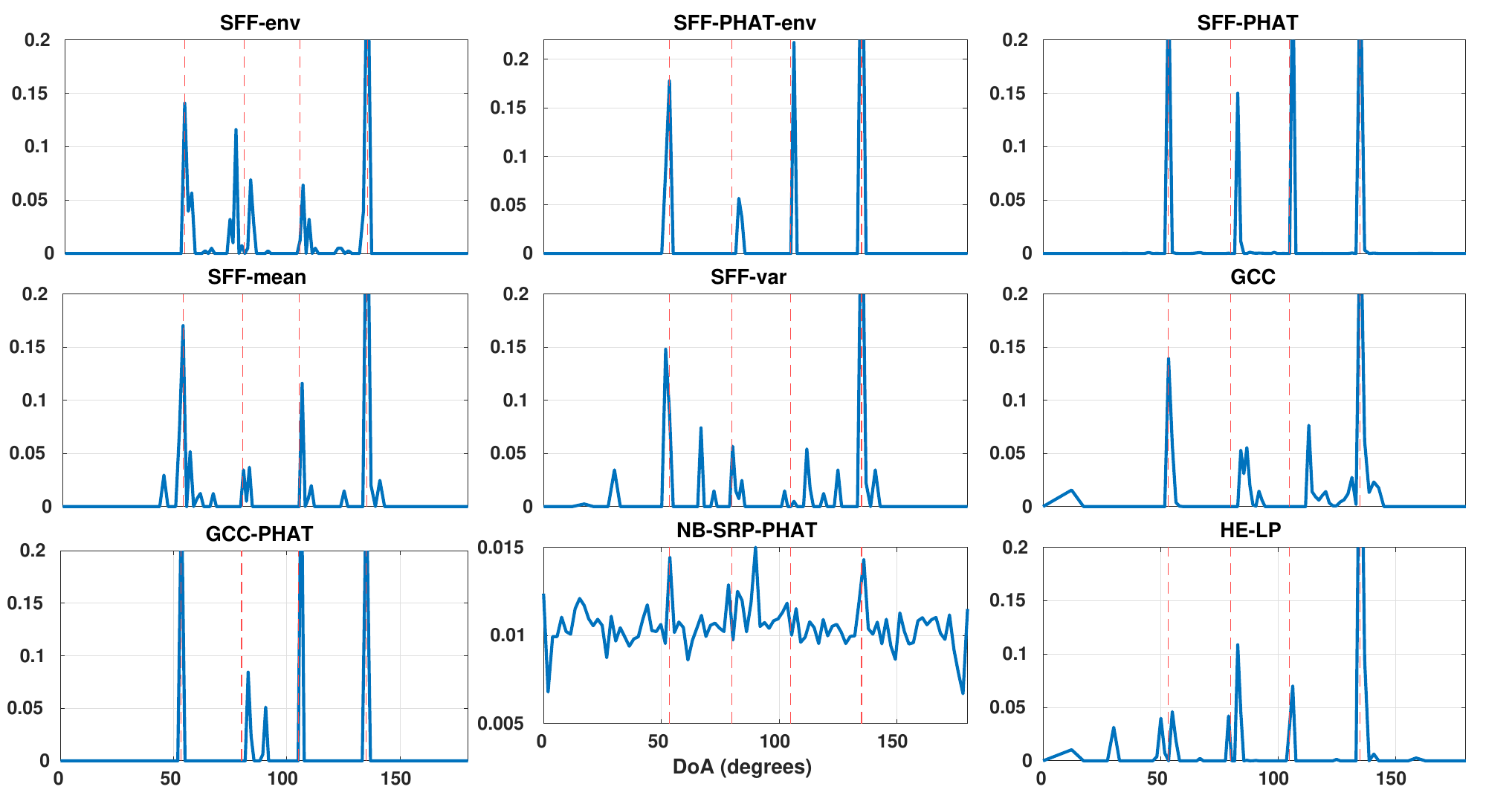}
  \caption[Histogram plots of various DoA estimators on a mixture from SiSEC]{Histogram plots of various DoA estimators on a mixture from SiSEC database with 4 female speakers and RT60 = 250~ms. The chosen DoA estimator is indicated as the title of the subplot. Y-axis of all the DoA estimators, except that of NB-SRP-PHAT, represents $\alpha$. The red dashed lines are the ground truth DoAs. SFF-PHAT is the best performing (with the highest $\alpha$) and is followed closely by GCC-PHAT and SFF-PHAT-env.}
  \label{fig:SiSEC_fem4_250ms}
\end{figure}

\begin{figure}[h!]
  \centering
  \includegraphics[scale=0.45]{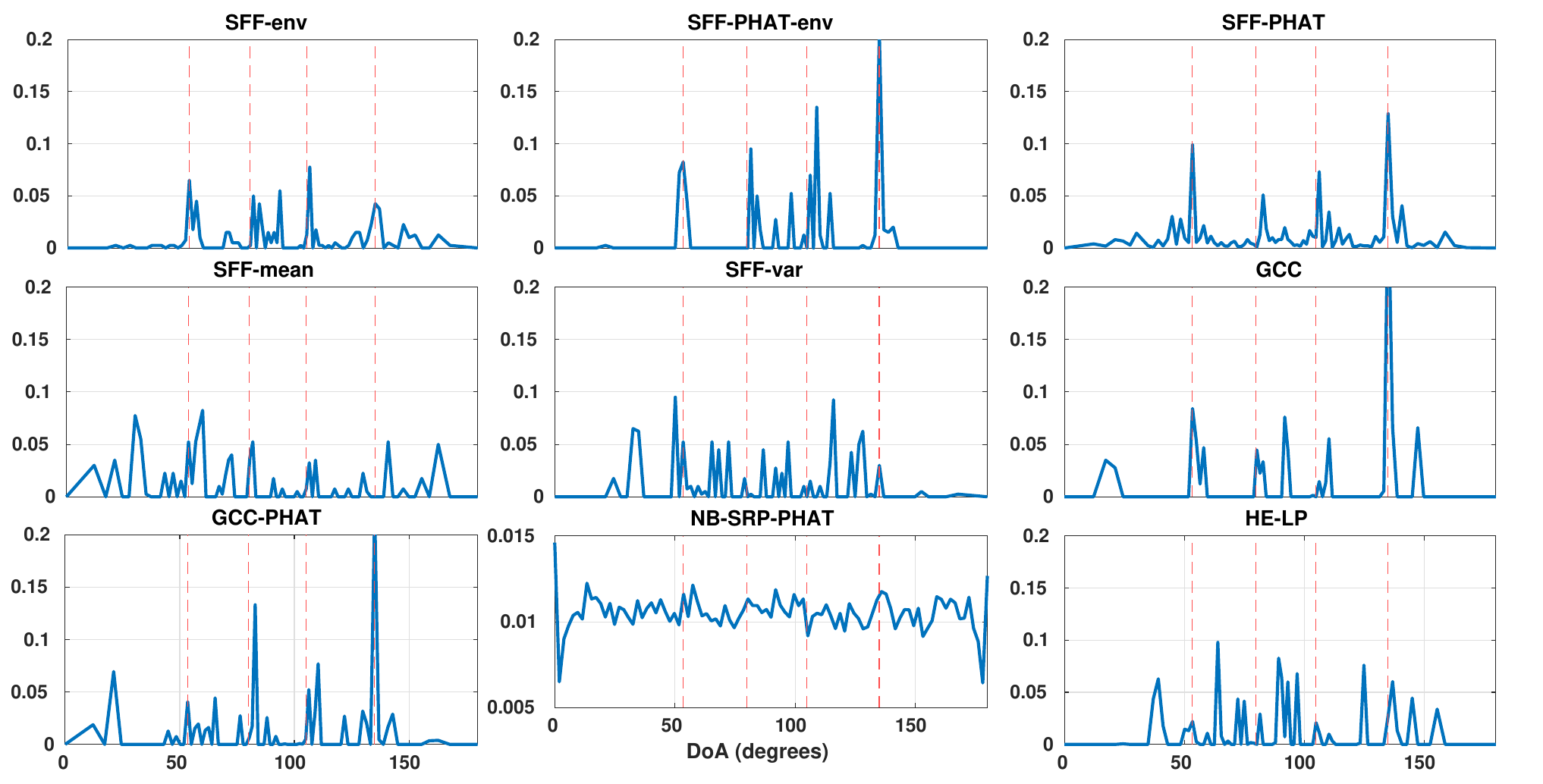}
  \caption[Histogram plots of various DoA estimators on a mixture from SiSEC  corrupted with babble noise at SNR = 0dB.]{Histogram plots of various DoA estimators on a mixture from SiSEC database with 4 female speakers and RT60 = 250~ms, corrupted with babble noise at SNR = 0dB. The chosen DoA estimator is indicated as the title of the subplot.  Y-axis of all the DoA estimators, except that of NB-SRP-PHAT, represents $\alpha$. The red dashed lines are the ground truth DoAs. SFF-PHAT and SFF-PHAT-env are the best performing (all the speakers are detected, least MAE and MAEfine and high $\alpha$). In GCC-PHAT, the source at 50 degrees is undetected due to a spurious peak at around 20 degrees.}
  \label{fig:SiSEC_fem4_250ms_noisy}
\end{figure}

\indent To visualize the performance differences among various DoA estimators, the histogram plots obtained on a recording from the SiSEC database are plotted in Figure~\ref{fig:SiSEC_fem4_250ms}. The recording chosen is that of 4 concurrent female speakers in a room with a reverberation time of 250~ms. It is evident from the figure that most of the DoA estimators can detect multiple speakers because SiSEC data is with less reverberation (up to 250ms). Figure~\ref{fig:SiSEC_fem4_250ms_noisy} shows the histogram plots  on noisy SiSEC recording obtained by adding babble noise at SNR = $0~dB$. As seen from the figure, SFF-PHAT and SFF-PHAT-env are the best DoA estimators with all the sources being correctly detected with less azimuth errors and good $\alpha$. In GCC-PHAT ,the source at 50~degrees is undetected due to spurious peak at around 20~degrees.        

\clearpage
\subsection{\label{subsec:SNR_wt_doa_srp_nmf}Weighted DoA Estimators}DoA estimators are susceptible to strong reverberation and noise. Many weighting functions have been proposed to emphasize the source dominant TF bins and to de-emphasize noise or reverberation dominant bins in order to improve the robustness of the DoA estimators as follows:
\begin{equation}
   \label{eq:mask_gcc_phat}
    \mathcal{J}_{\text{mask}}[\tau(\theta),b] =   \sum_{k}{\mathcal{\hat{M}}(k,b)\frac{X_{1}[k,b]X^*_{2}[k,b]}{|X_{1}[k,b]||X^*_{2}[k,b]|}e^{-j2\pi f_k \tau(\theta)}},
\end{equation}
where $\mathcal{\hat{M}}(k,b)$ represents the importance of $(k,b)^{th}$ bin for DoA estimation. High values represent source dominant bins, and low values correspond to noise or reverberation dominant bins. \\ \par\indent Several frequency-dependent weighting techniques are proposed for GCC-PHAT. In \cite{SNRwt_2003, 2003_Roman_Dwang_loc, BLANDIN20121950, 2007_MaskforSSL_Jean, 2015_MaskForSSL_Grondin}, SNR-based weightings are used, where SNR is computed using either rule-based methods or voice activity detection based algorithms or inter-channel coherence methods. However, the SNR weighting approach assumes stationary noise and fails in non-stationary noise scenarios. Another approach to address this problem is by summing only the bins dominated by the direct path of the target source in the GCC-PHAT optimization function, where the direct-path dominant TF bins may be identified by deep learning \cite{2019MaskGccPhat}. In a recent contribution, \cite{DBLP:journals/ejasmp/ThakallapalliGM21}, Non-negative Matrix Factorization (NMF)-based weighting is used to highlight single source dominant bins.\\ \par\indent As seen in Section~\ref{subsub:noisySisec}, SFF-PHAT outperforms GCC-PHAT and SFF-PHAT-env is comparable to GCC-PHAT in the noisy SiSEC data. In this section, we test the robustness of the best performing SFF-based estimators by comparing them with an improved GCC-PHAT, which is obtained by NMF weighting. The NMF weighting as proposed in \cite{DBLP:journals/ejasmp/ThakallapalliGM21} is used in conjunction with GCC-PHAT(referred to as GCC-NMF), in which only the source dominant TF bins contribute towards DoA estimation. In Table~\ref{tab:gcc_nmf}, the performance of GCC-NMF is compared with the best performing SFF- and GCC- based methods in terms of F-measure, MAE, and MAEfine. $\alpha$ is not reported because in GCC-NMF, as in the NB-SRP-PHAT, a single source can be detected multiple times in the same frame as described in Section~\ref{subsub:sisec&locata}. The parameters in GCC-NMF are set as follows: window size of 8000 samples, hop size = 80 samples, dictionary size =55, and the penalty term = 60. The window size is the same as that used for other DoA estimators. The DFT points are equal to the window size. After obtaining the STFT spectrogram, 512 discrete frequencies (as used in the other DoA estimators) equally spaced between 0 and 8000~Hz are chosen for further operations.


\begin{table}[t]
\centering

\caption [Performance comparison of SFF-based and GCC-NMF DoA estimators on noisy SiSEC.] {\label{tab:gcc_nmf} {Performance comparison of the best performing SFF-based (SFF-\textit{PHAT-env}, SFF-PHAT) with GCC-PHAT and GCC-NMF DoA estimators in terms of F-measure, MAE, and MAEfine on noisy SiSEC data. The noises (white, babble, pink, volvo and machine gun (M. gun)) are added at SNR = 0~dB. The best metrics amon the various DoA estimators are highlighted in \textbf{bold}.}}

\begin{tabular}{|l||l|l|l|l|}

\hline
& \multicolumn{2}{c|}{{SFF - based}}                                                                                                                                                                    & \multicolumn{1}{c|}{{Baseline}}&{Weighted }\\   

 & \multicolumn{2}{c|}{{ DoA estimators}}  & \multicolumn{1}{c|}{{ DoA estimators}}&{GCC-PHAT} \\  
 \hline
 &  {SFF-PHAT-env} &  {SFF-PHAT} & {GCC-PHAT} &{GCC-NMF} \\
 
 
  \hline
 \multicolumn{5}{|c|}{\textbf{SiSEC+White}}                                                                                                                                                                                                                                                                                                                                                                                                                                                                                                                        \\ \hline

 F-measure  & 0.5&{0.69}&{0.57}&\textbf{0.81}\\
 \hline
 MAE &{16.28}&\textbf{10.54}&{11.76}&12.71 \\
 \hline
 MAEfine &{1.98}&\textbf{1.62}&1.79&2.38\\
 \hline
 \multicolumn{5}{|c|}{\textbf{SiSEC+Babble}}                                                                                                                                                                                                                                                                                                                                                                                                                                                                                                                                                                                                                      \\ \hline
 F-measure &{0.91}&\textbf{0.93}&0.740&{0.92}\\
 \hline
 MAE &{3.45}&\textbf{1.14}&7.47&4.42\\
 \hline
 MAEfine&1.62&\textbf{0.79}&{0.93}&1.62\\ 

   \hline
 \multicolumn{5}{|c|}{\textbf{SiSEC+Pink}}                                                                                                                                                                                                                                                                                                                                                                                                                                                                                                                                                                                                                      \\ \hline
 F-measure &{0.54}&{0.74}&{0.60}&\textbf0.78\\ 
 \hline
 MAE&{11.34}&\textbf{8.03}&{11.37}&10.18\\ 
 \hline
 MAEfine &1.75&\textbf{1.25}&1.80&1.8\\
 \hline

 \multicolumn{5}{|c|}{\textbf{SiSEC+Volvo}}                                                                                                                                                                                                                                                                                                                                                                                                                                                                                                                                                                                                                      \\ \hline
 F-measure&{0.97} &\textbf{1.00}  &0.90 &\textbf{1.00}\\
 \hline
 MAE&0.95 &\textbf{0.78}  &4.16&1.12 \\
 \hline
 MAEfine &0.95 &\textbf{0.78} &{0.84} &1.12\\
 \hline
 \multicolumn{5}{|c|}{\textbf{SiSEC+M.Gun}}                                                                                                                                                                                                                                                                                                                                                                                                                                                                                                                                                                                                                      \\ \hline
 F-measure &0.97 &\textbf{1.00} &0.86 &\textbf{0.99}\\
 \hline
 MAE &{1.03}&\textbf{0.78}&4.04&1.14\\
 \hline
 MAEfine  &1.03 &\textbf{0.78} &{0.85} &1.03 \\
 \hline
\end{tabular}
\end{table} 
From the table, it is observed that the performance of GCC-NMF is significantly better than its \textit{unweighted} counterpart the GCC-PHAT: F-measure is much higher, MAE is lower or comparable. The difference in MAEfine between the two methods is less than one degree. Further, SFF-PHAT has the lowest MAE and MAEfine in all the noises. F-measure of SFF-PHAT is slightly less in white noise but comparable in all the other noises. To summarize, SFF-PHAT seems to be performing better than GCC-NMF in most of the noises, thus proving the robustness of the SFF-PHAT DoA estimator.

\section{\label{sec:conclusions}Summary and Conclusions}
In this study, nine DoA estimators (one estimator exploiting excitation source feature in the time domain, three estimators exploiting spectral features in STFT  representation, four estimators exploiting excitation source features in SFF  representation, and one using spectral features in SFF representation) are evaluated in terms of detection (F-measure and percentage of frames in which speakers are detected ($\alpha$)) and accuracy metrics (gross and fine mean azimuth errors). Publicly available SiSEC and LOCATA data sets are used. Tests are conducted in high reverberation and low SNR conditions. The results indicate that, in SiSEC and LOCATA datasets, SFF-PHAT and GCC-PHAT are the best performing estimators with comparable performance. The proposed SFF-PHAT-env is the next best. In noise added SiSEC data, SFF-PHAT performs better than all the estimators and outperforms the GCC-PHAT. SFF-PHAT-env and GCC-PHAT have comparable performances. GCC operation in the SFF domain gives more robust performance than the GCC operation in the STFT domain, as evident by comparing SFF-PHAT with GCC-PHAT. In addition, in the SFF domain the spectral features seem more robust to degradations than the temporal features as evident by comparing SFF-PHAT (exploits spectral features) with SFF-PHAT-env (exploits temporal features).  Further, we show that the SFF-PHAT has lower MAE than the \textit{weighted} GCC-PHAT in all the noises and further, is comparable to the \textit{weighted} GCC-PHAT in other metrics, proving the robustness of DoA estimation using SFF-PHAT.







\bibliography{mybib}
\end{document}